\documentclass[prb,twocolumn,groupedaddress,showpacs]{revtex4}
\usepackage{graphicx}
\usepackage{subfigure}
\usepackage{amsmath,amssymb}
\def\lb{[}
\def\rb{]}

\begin{document}

\title{Numerical study of the random field Ising model at zero and positive temperature}

\author{Yong Wu}
\affiliation{Department of Physics, Virginia Tech, Blacksburg, VA 24061-0435}
\author{Jonathan Machta}
\affiliation{Department of Physics, University of Massachusetts, Amherst, MA 01003-3720}

\date{\today}

\begin{abstract}
In this paper the three dimensional random field Ising model is studied at both zero temperature and positive
temperature. Critical exponents are extracted at zero temperature by finite size scaling analysis of large
discontinuities in the bond energy. The heat capacity exponent $\alpha$ is found to be near  zero. The ground
states are determined for a range of external field and disorder strength near the zero temperature critical
point and the scaling of ground state tilings of the field-disorder plane is discussed. At positive temperature
the specific heat and the susceptibility are obtained using the Wang-Landau algorithm. It is found that sharp
peaks are present in these physical quantities for some realizations of systems sized $16^3$ and larger. These
sharp peaks result from flipping large domains and correspond to large discontinuities in ground state bond
energies. Finally, zero temperature and positive temperature spin configurations near the critical line are
found to be highly correlated suggesting a strong version of the zero temperature fixed point hypothesis.
\end{abstract}


\maketitle

\section{Introduction}
The random field Ising model (RFIM) is among the simplest non-trivial spin models with quenched disorder. It has
been intensively studied theoretically, experimentally, and in computer simulations during the last thirty years
but is still not well understood. Following the seminal discussion of Imry and Ma~\cite{ImMa75} it has been
proved that the RFIM has an ordered phase at low temperature and weak disorder when the dimension is greater
than two~\cite{GrMa82,Imbrie84,BrKu87}. It is generally believed that the transition from the ordered phase to
the disordered phase of the RFIM is continuous and is controlled by a zero temperature fixed
point~\cite{BrMo,Villain85,Fish86}. Since random field fluctuations dominate over thermal fluctuations at the
transition, the hyperscaling relation is modified as $(d-\theta)\nu = 2-\alpha$, where  $\theta$ is the
violation of hyperscaling exponent~\cite{Villain85,Fish86}.

\begin{figure}
\begin{center}
\includegraphics[width=2.5in]{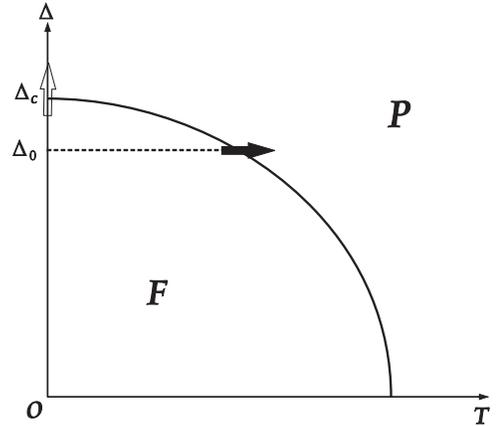}
\caption{Phase diagram of the RFIM. We study two types of phase transitions going from the ferromagnetic phase
(F) to the paramagnetic phase (P): The zero temperature transition (open arrow) occurs at $T=0$ and
$\Delta=\Delta_c$, and positive temperature transitions (solid arrow) occurs at a fix disorder $\Delta =
\Delta_0<\Delta_c$ and $T>0$.} \label{fig:phtr}
\end{center}
\end{figure}

The phase diagram of the RFIM is sketched in Fig.\ \ref{fig:phtr}. Phase transitions can occur from the
ferromagnetic phase (F) to the paramagnetic phase (P) at either zero temperature as a function of disorder
strength $\Delta$ at $\Delta = \Delta_c$, or as a function of temperature $T$ if disorder is fixed at
$\Delta_0<\Delta_c$. According to the zero temperature fixed point hypothesis, the zero temperature transition
and the positive temperature transitions belong to the same universality class. In this paper we use numerical
methods to study both kinds of transitions and connections between them.  One of our primary results is that,
for each realization of disorder, there is a strong correlation between ground state configurations near
$\Delta_c$ and thermal states near $T_c$ for $\Delta_0<\Delta_c$.

Currently, there is no controlled renormalization group analysis of the RFIM phase transition and Monte Carlo
simulations of the RFIM at positive temperature~\cite{RiYo,Rieg95,NeBa,MaNeCh00} are limited to small systems
because of very long equilibration times~\cite{Villain85,Fish86}. According to the zero temperature fixed point
hypothesis, many properties of the RFIM phase transition, including the values of critical exponents, can be
determined by studying the RFIM at zero temperature. The ground state of the RFIM can be found in polynomial
time~\cite{Ogielski} by efficient combinatorial algorithms so that zero temperature simulations are much faster
and allow for much larger system sizes than positive temperature simulations. Critical exponents have been
obtained from zero temperature studies~\cite{HaYo01,MiFi02,DuMa03} that are mostly consistent with the scaling
theories~\cite{BrMo,Villain85,Fish86}, series methods~\cite{gof} and real space renormalization group
approaches~\cite{CaMa,FaBeMc,EfSc}.

Much work has been done in determining the critical exponents, and the values of many exponents are well
established. However, the value of the heat capacity exponent $\alpha$ is still controversial. A recent zero
temperature study by Hartmann and Young~\cite{HaYo01} found $\alpha\approx-0.6$ for the three dimensional
Gaussian RFIM. The modified hyperscaling relation, however, predicts that $\alpha=2-(d-\theta)\nu\approx 0$,
given the well-accepted values $\theta\approx1.5$ and $\nu\approx1.1-1.4$. Therefore the quite negative value
found in Ref.\ \onlinecite{HaYo01} is inconsistent with the modified hyperscaling relation. Some older work at
zero temperature~\cite{NoUsEs} and Monte Carlo simulations~\cite{Rieg95} also found $\alpha$ quite negative. On
the other hand, Middleton and Fisher~\cite{MiFi02} also studied the  three dimensional Gaussian RFIM at $T=0$
and found $\alpha\approx-0.1$,  in agreement with the modified hyperscaling relation.

Dukovski and Machta~\cite{DuMa03} studied the ground states of the RFIM in the presence of an external field
$H$. They located a ``finite size critical point" for each realization of disorder, identified as the point of
degeneracy of three ground states in the the $H-\Delta$ plane with the largest discontinuity in magnetization.
They extracted critical exponents via finite size scaling of the discontinuities at that point. The reason to
focus on the finite size critical point was that this point can be regarded as the most singular point on the
$H-\Delta$ plane, and working at this point may reduce the influence of the regular part of physical quantities.
The value of heat capacity exponent they found was $\alpha \approx 0$, however, their results were less accurate
than those of Refs.\ \onlinecite{HaYo01} and \onlinecite{MiFi02}  because of the large amount of computational
work needed to locate the finite size critical point.

The work reported in this paper combines both zero temperature and positive temperature studies.  The zero
temperature studies extend the work of Ref.\ \onlinecite{DuMa03} in two directions. First, for each realization
of disorder we study points along the $H = 0$ line with large discontinuities in bond energy or magnetization to
determine the critical exponents. Finding discontinuities along the $H=0$ line requires much less computational
work than finding the finite size critical point while still adhering to the idea introduced in Ref.\
\onlinecite{DuMa03} of extracting critical exponents from the large discontinuities in each realization of
disorder.  We also find ground state spin configurations near these large discontinuities and compare them to
thermal states near positive temperature critical points. Second, we study the full set of ground states of the
RFIM in the critical region of the $H-\Delta$ plane and discuss the properties of the resulting ground state
tilings of this plane.

Since conventional Monte Carlo methods are not efficient for the study of the RFIM, we apply the Wang-Landau
algorithm~\cite{WaLa01} to the RFIM, which enables us to obtain the specific heat and the susceptibility over a
broad range of temperature with system size up to $32^3$. We find that some realizations display sharp peaks in
the specific heat and susceptibility. Inspired by the zero temperature fixed point hypothesis, we relate these
sharp peaks to the large discontinuities at zero temperature. We further study the thermal states (average spin
configurations) near the transition using the Metropolis algorithm and compare them to the ground states near
the zero temperature transition.  Some of this work has been previously announced in Ref.\ \onlinecite{WuMa05}.

In this paper we consider the  three dimensional RFIM with Gaussian random fields described by the Hamiltonian,
\begin{equation} \label{eq:hRFIM} {\mathcal{H}}=-\sum_{\langle
i,j\rangle}s_{i}s_{j}-\Delta\sum_{i}h_{i}s_{i}-H\sum_{i}s_{i},
\end{equation}where $H$ is the uniform external field, $\langle i,j\rangle$ indicates a sum over
all nearest neighbor sites $i$ and $j$ on a simple cubic lattice of linear size $L$ with periodic boundary
conditions.  The random fields $h_i$ are Gaussian random variables with mean zero and standard deviation one and
the strength of disorder is $\Delta$. The normalized fields $\{h_i\}$ define a realization of disorder and, for
a given realization of disorder we explore spin configurations and physical properties as a function of $H$, $T$
and $\Delta$.  Some of the physical quantities of interest include the magnetization $m$, defined
as\begin{equation} \label{eq:defm} m=\frac{1}{L^3}\sum_is_i,\end{equation}and the bond energy
$e$,\begin{equation} \label{eq:defe} e=-\frac{1}{L^3}\sum_{\langle i,j\rangle}s_is_j.\end{equation}

In the next section we discuss the scaling properties of large discontinuities in the bond energy at zero
temperature and use numerical results for these discontinuities to extract critical exponents and the critical
disorder strength.    In Sec.\ \ref{sec:gndpix} we obtain ground state portraits for the RFIM and discuss their
scaling properties.  Section \ref{sec:postemp} presents results of positive temperature simulations and, in
Sec.\ \ref{sec:gndtherm}, we discuss correlations between ground states and thermal states.  The paper closes
with a summary and discussion.

\section{Critical exponents at zero temperature}
\label{sec:measure}

At zero temperature, the problem of finding the ground state of the RFIM can be mapped to the MAX-FLOW problem
in graph theory, which is solvable in polynomial time~\cite{Ogielski}. We use a modified version of the
push-relabel algorithm \footnote{The algorithm is available from Andrew Goldberg's Network Optimization Library,
http://www.avglab.com/andrew/soft.html} to calculate the ground states~\cite{GoTa88,GoCh97}.

\begin{figure}\includegraphics[width=3in]{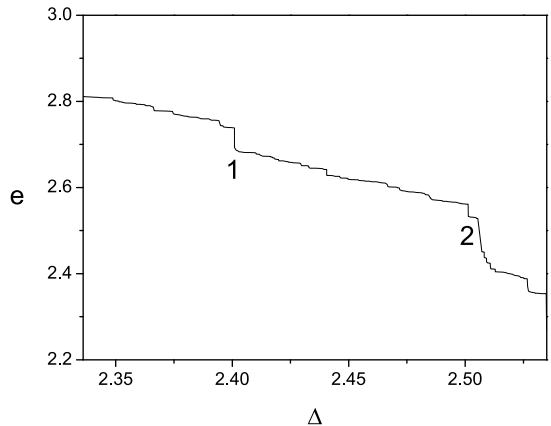}
\caption{Bond energy as a function of disorder strength $\Delta$ for a single realization of disorder (seed
1003).  Numbers 1 and 2 indicate the two biggest jumps in the bond energy. } \label{fig:ejumps}
\end{figure}

\begin{figure}\includegraphics[width=3in]{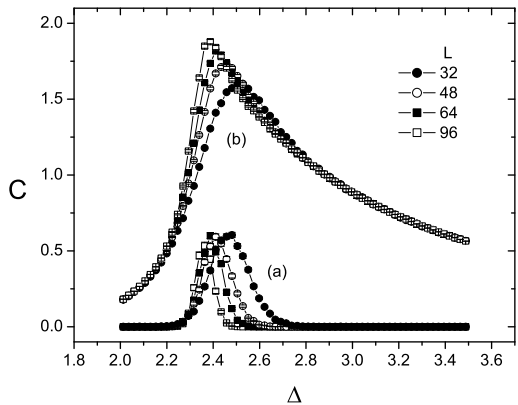}
\caption{ Two components of the specific heat. (a) The contribution to the specific heat from the largest $N(L)$
bond energy jumps, where $N(L)$ is the number of disorder realizations for each system size $L$. (b) The
contribution from all other bond energy jumps. } \label{fig:minus1}
\end{figure}

The specific heat at $T=0$ is defined~\cite{HaYo01} as
\begin{equation} \label{eq:defc} C=\frac{\partial \lb
e\rb}{\partial\Delta},
\end{equation}where $e$ is the bond energy defined in Eq.\ (\ref{eq:defe}) and the square brackets denote averaging over disorder realizations.
At zero temperature, for each realization of normalized random fields the bond energy changes discontinuously as
a function of the strength of disorder $\Delta$. An example of a single realization is shown in Fig.\
\ref{fig:ejumps} and illustrates the point that the sizes of the discontinuities vary widely. In Refs.\
\onlinecite{HaYo01} and \onlinecite{MiFi02} all bond energy jumps are included in the calculation of the
specific heat exponent. However, small jumps are presumably part of the analytic background rather than the
singular behavior. The smallest bond energy jump, for example, is $4$ for all system sizes. We therefore analyze
large jumps in the bond energy to focus on the critical singularity. From a sample of $N(L)$ disorder
realizations of systems of size $L$, let $\lb \delta e_1 \rb$ be the average over the $N(L)$ largest bond energy
jumps and let $\lb \Delta_1 \rb$ be the average of the disorder strength at these jumps.  Each realization of
disorder  typically contributes one bond energy jump and one disorder strength to these averages though some
realizations contribute nothing and some several values.  Figure \ref{fig:minus1} shows the specific heat
decomposed into two components, component (a) is due to the largest $N(L)$ jumps while component (b) arises from
all other jumps. One can see that the large jumps make a significant contribution to the full specific heat and
we will use this component to extract critical exponents. The finite size scaling of the specific heat is
expected to obey
\begin{equation}
C \sim L^{\alpha/\nu} \tilde{C}((\Delta-\Delta_c)L^{1/\nu}) \label{eq:fssh}
\end{equation}
Though the two components of the specific heat shown in Fig.\ \ref{fig:minus1} obviously behave differently from
one another,   our primary assumption is that the finite-size scaling of the full specific heat also applies to
the component of the specific heat from the largest bond energy jumps.  Indeed we believe that the large jumps
provide better data to obtain critical exponents from small systems than the full specific heat because this
component is undiluted by the analytic background.  We discuss more detailed scaling assumptions about large
jumps later in this section.  Note that the peak height for component (a) barely changes with system size
suggesting that the specific heat exponent $\alpha$ is near zero.

Integrating Eq.\ (\ref{eq:fssh}) as applied to the component from the largest discontinuities, we obtain a
finite size scaling ansatz for the large jumps,
\begin{equation} \label{eq:sce}  \lb \delta e_1 \rb \sim L^{(1-\alpha)/\nu} .
\end{equation}
where $\alpha$ is the specific heat exponent and $\nu$ is the correlation length exponent.    Table
\ref{tab:raw} gives the average size of the large bond energy jumps as a function of system size $L$. A fit of
the form given in Eq.\ (\ref{eq:sce}) yields $(1-\alpha)/\nu=0.842\pm0.003$ with goodness of fit parameter
$Q\approx0.7$ ($Q\equiv\Gamma(d/2,\chi^2/2)$ with $d$ the number of degrees of freedom and $\Gamma$ the
incomplete gamma function).

The displacement of the average position of the large jumps from the infinite volume limit and the standard
deviation of the positions of the large jumps are each measures of the width of the critical region and,
following Ref.\ \onlinecite{AhHa}, we assume that they satisfy the finite size scaling relations,
\begin{equation} \label{eq:scd} \lb\Delta_1\rb - \Delta_c \sim
L^{-1/\nu},
\end{equation}
and
\begin{equation}
\label{eq:scdsq} \sqrt{\lb (\Delta_1-\lb \Delta_1\rb)^2\rb}\sim L^{-1/\nu}.
\end{equation}
where $\nu$ is the correlation length exponent and $\Delta_c$ is the infinite size critical disorder strength.
Table
 \ref{tab:raw} gives the standard deviation of the position of the largest jump and a fit to Eq.\ (\ref{eq:scdsq}) yields $1/\nu=0.79\pm0.01$ with $Q\approx0.4$.
Finally, Table  \ref{tab:raw} gives $\lb\Delta_1\rb$ and, using the previously obtained value, $1/\nu=0.79$ a
fit to Eq.\ (\ref{eq:scd}) yields $\Delta_{c}=2.280\pm0.003$ with $Q\approx0.2$.

\begin{table}
\centering \caption{Data from ground state simulations for the average largest jump, and the average and
standard deviation of the disorder strength at the largest jumps as a function of system size $L$.}
\begin{tabular*}{0.48\textwidth}{@{\extracolsep{\fill}}lccc}
\hline\hline $L$&$\lb \delta e_1 \rb$&$\lb (\Delta_1-\lb \Delta_1\rb)^2\rb^{1/2}$&$\lb\Delta_1\rb$\\
\hline
32&0.1208(4)&0.0904(5)&2.4915(10)\\
48&0.0857(5)&0.0666(6)&2.4326(10)\\
64&0.0675(5)&0.0524(5)&2.4017(10)\\
96&0.0480(6)&0.0385(7)&2.3709(10)\\
\hline\hline
\end{tabular*}
\label{tab:raw}
\end{table}

The second and third largest jumps also presumably reflect the critical singularity. We repeated the foregoing
calculations  for the largest $kN(L)$ jumps, where our data allow us to go up to $k=3$. The results are listed
in Table\ \ref{tab:result}. We arrive at the following best estimates of the critical exponent and the infinite
volume critical disorder,
\begin{align}\label{eq:result} \frac{1-\alpha}{\nu} =
0.842\pm0.004, \quad  \nu = 1.25 \pm 0.02 \nonumber \\  \Delta_c = 2.282 \pm 0.002, \quad \alpha =
-0.05\pm0.02.\end{align} where the error bars include statistical errors from all three $k$ values.

\begin{table}
\caption{Critical exponents extracted from the largest $kN(L)$ jumps, where $N(L)$ is the number of realizations
for system size $L$.  Errors are purely statistical.}\centering
\begin{tabular*}{0.48\textwidth}{@{\extracolsep{\fill}} l c c c}
\hline \hline
$k$&$(1-\alpha)/\nu$&$1/\nu$&$\Delta_c$\\
\hline
1&0.841(4)&0.79(1)&2.282(2)\\
2&0.842(4)&0.80(1)&2.282(2)\\
3&0.844(3)&0.81(1)&2.283(1)\\
\hline \hline
\end{tabular*}
\label{tab:result}
\end{table}

Our values of the $(1-\alpha)/\nu$ and $\Delta_c$ are consistent with some previous calculations and $\alpha$ is
found to be near zero, which is in agreement with Ref.\ \onlinecite{MiFi02}. But the value of $\nu$ we have
calculated is smaller than recent results quoted in Refs.\  \onlinecite{HaYo01} and \onlinecite{MiFi02}. In
Table \ref{tab:comp} our calculated values of the exponents are listed in comparison with some recent work. Our
values of $(1-\alpha)/\nu$ and $1/\nu$ gives $(2-\alpha)/\nu \approx 1.64$. Applying the modified hyperscaling
relation and the inequality $\theta \ge d/2-\beta/\nu$ ~\cite{Fish86,ScSo85}, one has $\beta/\nu \ge 0.14$,
which is inconsistent with other work. We believe that our value of $(1-\alpha)/\nu$ is more reliable than our
value of $1/\nu$. The fit for $1/\nu$ starts from size $L=32$ and would be quite poor if the $L=16$ data were
included suggesting significant finite size correction. On the other hand, if  the $L=16$ data were included,
the fit would still be good for $(1-\alpha)/\nu$ and there would be no change in the resulting value.

\begin{table}
\caption{A summary of recent estimates of $\Delta_c$, $\nu$, $(1-\alpha)/\nu$ and $\alpha$, either calculated by
ground state (GS) or Monte Carlo (MC) simulations.}\centering
\begin{tabular}{lccccc}
\hline\hline Ref.&$\Delta_c$&$\nu$&$(1-\alpha)/\nu$&$\alpha$&method\\
\hline
This work&2.282(2)&1.25(2)&0.842(4)&-0.05(2)&GS\\
~\onlinecite{DuMa03}&2.29(2)&1.1(1)&0.80(3)&0.12&GS\\
~\onlinecite{HaYo01}&2.28(1)&1.36(1)&1.20&-0.63(7)&GS\\
~\onlinecite{MiFi02}&2.270(4)&1.37(9)&0.82(2)\footnote{This value was calculated from scaling of the bond energy. They found $(1-\alpha)/\nu=0.74(2)$ by relating it to the fractal dimension of the surface of spin clusters.}&-0.12(12)&GS\\
~\onlinecite{AnSo}&2.26(1)&1.22(6)& & &GS\\
~\onlinecite{HaNo}&2.29(4)&1.19(8)& & &MC\\
~\onlinecite{NoUsEs}&2.37(5)&1.0(1)&1.55&-0.55(20)&MC\\
\hline\hline
\end{tabular}
\label{tab:comp}
\end{table}

Next we take a closer look at the finite size scaling properties of the distribution of discontinuities in the
bond energy. Bond energy jumps result from flipping domains. Considering $N(L)$ ($N\gg1$) realizations with
system size $L$, we assume that there is a renormalization group transformation mapping them to $N(L^{\prime})$
realizations with system size $L^{\prime}$, and the flipped domains are transformed such that the average bond
energy jump $\lb \delta e \rb$ and the average disorder where jumps occur $\lb \Delta \rb$ conform to the
already known scaling relations, Eqs.\ (\ref{eq:sce}) and (\ref{eq:scd}). In order to exclude small jump that do
not scale properly, we introduce a lower cut-off $\delta e_{\min}$ for bond energy jumps, which should scale the
same way as $ \lb \delta e \rb$,
\begin{equation}
\label{eq:scemin} \delta e_{\min} \sim L^{(1-\alpha)/\nu}. \end{equation} The total number of bond energy jumps
larger than the scaled lower cut-off is independent of the system size, since the jumps in systems with
different sizes are connected by the renormalization group transformation. Defining scaled variables $u=\delta e
L^{-(1-\alpha)/\nu}$ and $v=(\Delta - \Delta_c)L^{-1/\nu}$, the number of jumps occurring in a small
neighborhood of $(u,v)$ should also be invariant for different system sizes. It then follows that the
probability $P(\delta e, \Delta)$ of having a bond energy jump with size $\delta e$, $\delta e>\delta e_{\min}$
and position $\Delta$ is proportional to a given normalized probability distribution function $\tilde{P}(u,v)$,
\begin{equation} \label{eq:scp1} P(\delta e, \Delta) \propto \tilde{P}(\delta e
L^{(1-\alpha)/\nu}, (\Delta-\Delta_c)L^{1/\nu}).
\end{equation} The normalization of $P(\delta e, \Delta)$ then gives
\begin{equation}
\label{eq:scp2} P(\delta e, \Delta) = L^{(2-\alpha)/\nu} \tilde{P}(\delta e L^{(1-\alpha)/\nu},
(\Delta-\Delta_c)L^{1/\nu}).
\end{equation}
Letting $a = \delta e_{\min}L^{(1-\alpha)/\nu}$ and integrating gives the scaling of the specific heat due to
big jumps $C_b$,
\begin{eqnarray} \label{eq:scc} C_b & = & \int_{\delta e_{\min}}{\delta e P(\delta e, \Delta)} \,  d \delta e \nonumber \\
& \sim & L^{\alpha/\nu} \tilde{C}((\Delta-\Delta_c)L^{1/\nu}, a),
\end{eqnarray}where $\tilde{C}(v,a) =
\int_{a}{u\tilde{P}(u,v)\,du}$ is some scaling function. By integrating Eq.\ (\ref{eq:scp2}) over $\delta e$ we
derive the probability distribution of the disorder strength where big bond energy jumps occur,
\begin{equation} \label{eq:pd}
P_b(\Delta)= L^{1/\nu}\tilde{Q}((\Delta-\Delta_c)L^{1/\nu}, a),
\end{equation}
where $\tilde{Q}(v,a)= \int_{a}{\tilde{P}(u,v)\,du}.$  We then recover the scaling of the average disorder
strength given in Eq.\ (\ref{eq:scd}) The setup of a scaled lower cut-off should be equivalent to picking the
$kN(L)$ largest jumps for each size $L$, where $k$ is some fixed number, because the total number of jumps is
invariant under the renormalization group transformation. We test this hypothesis by fixing the constant
$a=\delta e_{\min} L^{(1-\alpha)/\nu}$ and count how many jumps larger than the lower cut-off there are for each
system size. The result is listed in Table\ \ref{tab:eqv}. One can see that if the scaled lower cut-off is not
too small, the number of bond energy jumps larger than the cut-off goes to a constant independent of the system
size $L$.

\begin{table}
\caption{Number of bond energy jumps larger than the scaled lower cut-off $\delta e_{min}$. The cut-off
satisfies that $a=\delta e_{\min}L^{(1-\alpha)/\nu}$ is a constant.}\centering
\begin{tabular*}{0.48\textwidth}{@{\extracolsep{\fill}} l c c c c c}
\hline\hline
&$L=16$&$L=32$&$L=48$&$L=64$&$L=96$\\
\hline
$a=0.05$& 69.3(1)& 75.1(1)& & &\\
$a=0.1$& 29.03(6)& 30.89(6)& & & \\
$a=0.6$& 3.16(1)& 3.04(1)& 3.01(1)& 2.99(2)& 3.00(3)\\
$a=1$& 1.585(5)& 1.533(7)& 1.53(1)& 1.52(1)& 1.54(2)\\
\hline\hline
\end{tabular*}
\label{tab:eqv}
\end{table}

\begin{figure}\includegraphics[width=3in]{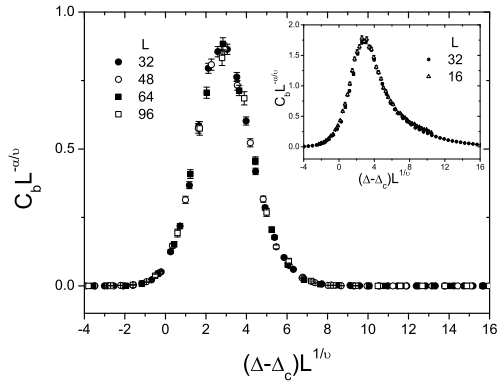}
\caption{Data collapse of the specific heat. The cut-off is $a=\delta e_{\min} L^{(1-\alpha)/\nu} = 1.0$. System
sizes range from $16^3$ to $96^3$. The inset shows data collapse of the specific heat for system size $16^3$ and
$32^3$, and the cut-off is $\delta e_{\min} L^{(1-\alpha)/\nu} = 0.1$.} \label{fig:collapse}\end{figure}

Figure \ref{fig:collapse} illustrates the data collapse of the specific heat predicted by Eq.\ (\ref{eq:scc})
for system sizes $32^3$, $48^3$, $64^3$ and $96^3$. In  conventional data collapse the scaling function
$\tilde{C}(x)$ behaves like $x^{-\alpha}$ as $x\rightarrow\infty$, but in Fig.\ \ref{fig:collapse} the tail of
the curve decays faster than a power law. The main plot in Fig.\ \ref{fig:collapse} has the bond energy jump
cut-off set as $a=\delta e_{\min}L^{(1-\alpha)/\nu}=1$.  Table \ref{tab:eqv} shows that for $a=1$ only a few
largest jumps per realization contribute to $C_b$ and, since these jumps are concentrated near the critical
point, we do not expect $C_b\sim(\Delta-\Delta_c)^{-\alpha}$. However, the tail should approach $x^{-\alpha}$,
if the number of jumps included is increased, or equivalently, $a$ is reduced.  The inset in Fig.\
\ref{fig:collapse} shows $C_b$ with a smaller cut-off, $a=\delta e_{\min}L^{(1-\alpha)/\nu}=0.1$.  With this
cut-off data is available for $16^3$ and $32^3$ systems only. The inset illustrates that, as the cut-off is
lowered, the tail of the scaling function expands and presumably approaches the asymptotic $x^{-\alpha}$ shape.

\section{Ground States Pictures and Scaling Relations}
\label{sec:gndpix}

The tiling of the $H-\Delta$ plane by ground states is the subject of this section. To study this tiling, we
find all ground states within a certain range of disorder $\Delta$ and external field $H$ in the critical
region. Since the Hamiltonian of the RFIM is linear with respect to the external field $H$ and the strength of
disorder $\Delta$, each spin configuration is represented by a plane in the $H-\Delta-\mathcal{H}$ coordinate
system. Ground states are spin configurations that are locally lowest and the set of all ground states form a
convex surface in this coordinate system.    Spin configurations are ground states within regions of $H$ and
$\Delta$ bounded by neighboring ground state planes so that a given spin configuration is the ground state
within a polygonal region. At boundaries of these polygons, and intersection points of boundaries, ground states
are degenerate. We are particularly interested in the degenerate points that are common points of three ground
states. We call these ``triple points."

The structure of the ground state energy surface can be visualized by projecting it onto the $H-\Delta$ plane
where it becomes a tiling of the plane by polygons.  The computational method for finding this tiling is closely
related to the method developed in Ref.\  \onlinecite{DuMa03}. In that work, the first order line, which is a
set of boundaries that separates the two ordered phases with positive and negative magnetization, respectively,
was followed and the ``finite size critical point" was identified by finding the triple point having the largest
discontinuity in magnetization. The finite size critical point was regarded as the most singular point, and
critical exponents were extracted via finite size scaling of magnetization and bond energy discontinuities at
the point. In this paper we use a method similar to the techniques used in Ref.\ \onlinecite{FrGoOrVi} and
\onlinecite{DuMa03} to map out all the ground states for any given realizations within a certain region on the
$H-\Delta$ plane near the finite size critical point.

Our algorithm performs a breadth-first search of ground states. The starting point of the search is the finite
size critical point located by the algorithm of Ref.\ \onlinecite{DuMa03}. For each point where more than two
ground states are degenerate we already have the ground states around the point and the coexistence lines
separating them (one locates the point by finding ground states around it).  We then follow the lines and search
for the next adjacent degenerate point using the following method. Starting from the given degenerate point $p$,
we extend the coexistence line separating ground states $P_1$ and $P_2$ with some preselected step size until we
meet a point $q_0$ on which the ground state $Q_0$ is different from both $P_1$ and $P_2$. The actual adjacent
degenerate point is typically passed over, because of the the fixed step size is too large. We then locate the
intersection point of $P_1$, $P_2$ and $Q_0$ and name it $q_1$ on which the ground state is $Q_1$. If $q_1=q_0$
then $q_0$ is obviously the point we want. Otherwise we find the intersection point of $P_1$, $P_2$ and $Q_1$
and name it $q_2$. The process can be repeated and the sequence $\lbrace q_n \rbrace$ will eventually converge
to the adjacent degenerate point due to the convexity of the ground state surface. The process of finding
adjacent degenerate points is iterated recursively until it reaches the outside of the predefined region, or it
finds a point that has already been visited. By connecting degenerate points with straight lines, all ground
states within the region are identified.

\begin{figure}
\begin{center}
\subfigure[]{\label{fig:allgs}\includegraphics[width=3in]{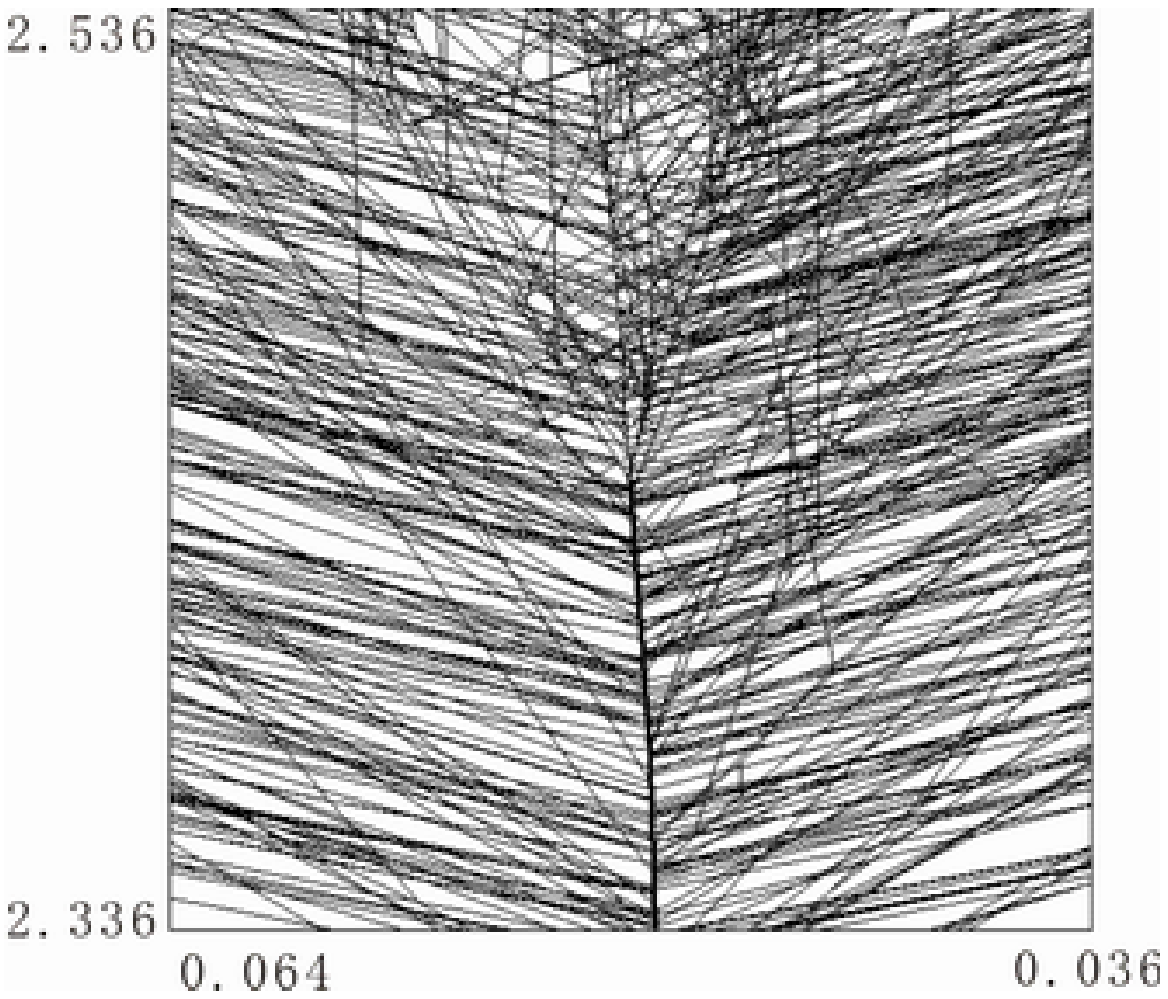}} \subfigure[]{\label{fig:simgs}\includegraphics[width=3in]{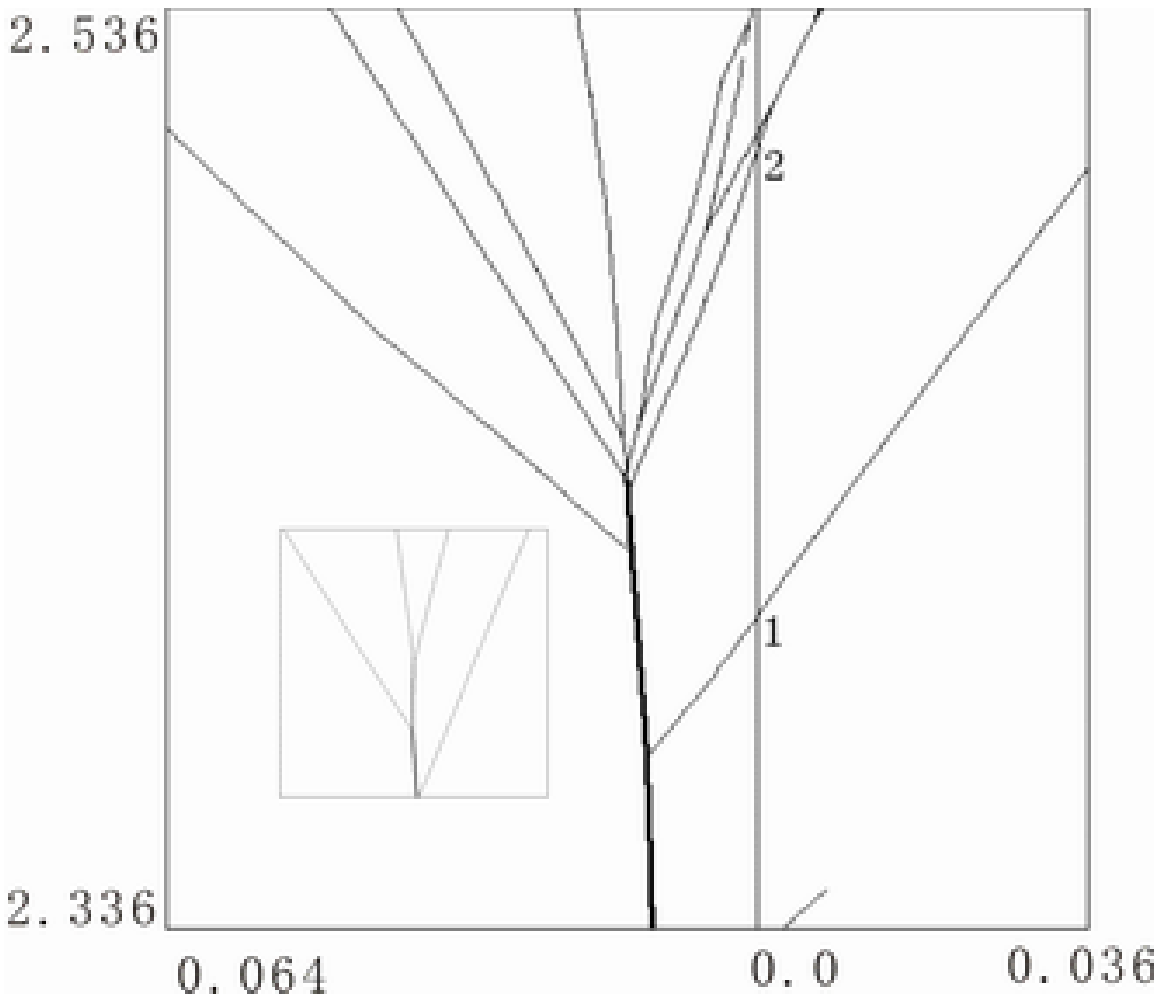}} \\
\includegraphics[width=0.6in]{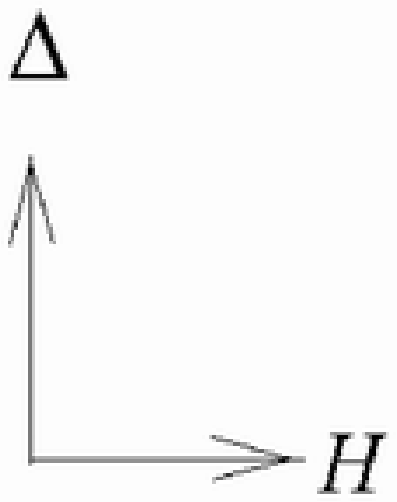}
\caption{Ground states of a given realization (seed 1003) with system size $32^3$ in the $H-\Delta$ plane. (a)
All the ground states of a single realization with $L=32$. The lines are coexistence lines of two ground states.
The thickness of a line is proportional to the magnetization jump across the line.(b) The same realization as in
(a), but only coexistence lines with bond energy jumps $\delta e>0.03$ are shown. Numbers 1 and 2 correspond to
the two largest jumps shown in Fig.\ \ref{fig:ejumps}.  The inset in (b) is a blow up of the region around the
triple point identified as the finite size critical point and also showing the triple points immediately above
and below it.} \label{fig:gndstates}
\end{center}
\end{figure}

\begin{figure}
\begin{center}
\subfigure[]{\label{fig:triple}\includegraphics[width=0.8in]{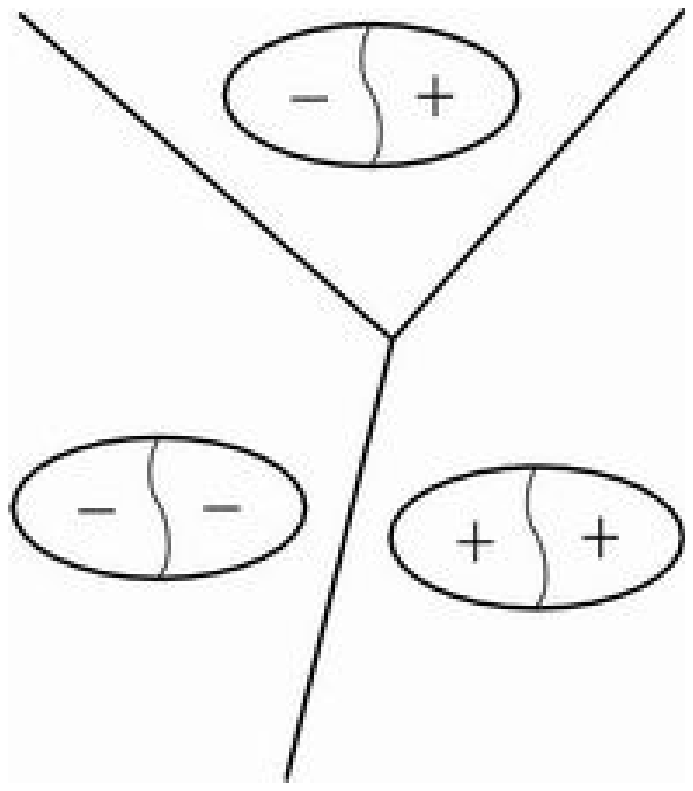}} \hspace{0.4in}
\subfigure[]{\label{fig:quadr}\includegraphics[width=1.2in]{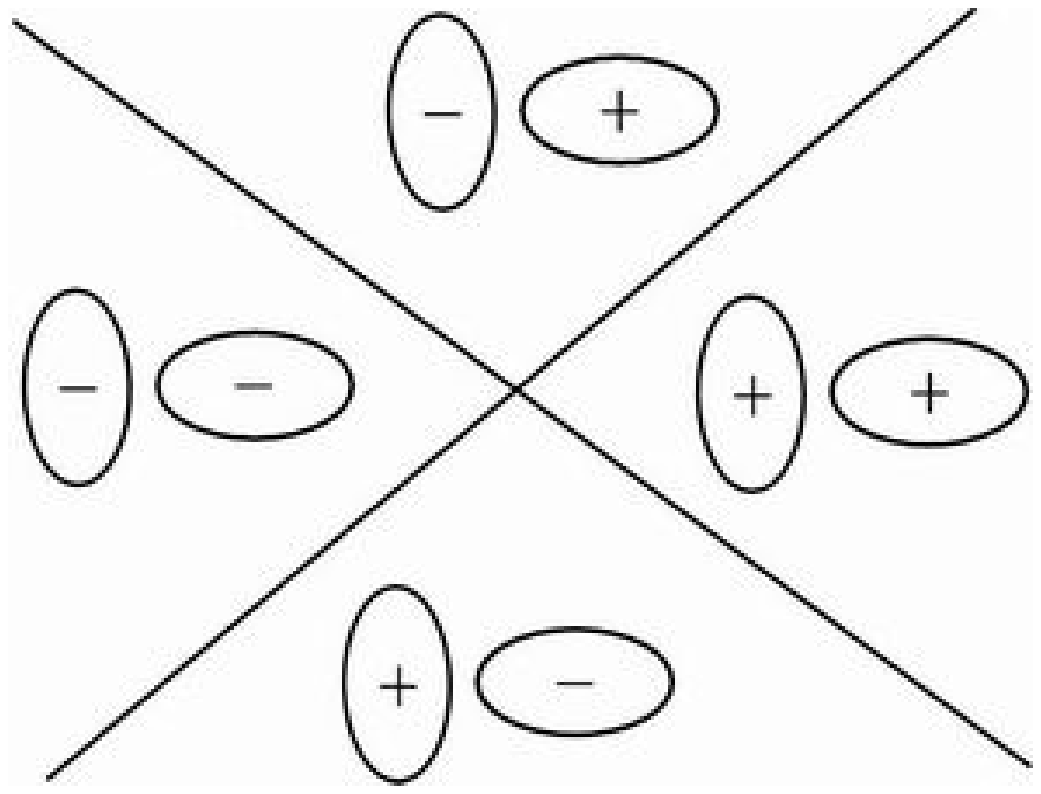}}
\end{center}
\caption{ (a) Three states are degenerate at a triple point. The ``$+$'' and ``$-$" sign are used to indicate
the direction of spins in a domain. The spins in the domain are all pointing up (denoted as ``$+ +$") or all
down (``$- -$") in the ground states below the triple point, while the domain breaks up (``$+ -$") in the ground
state on top of the triple point. (b) Four states separated by two intersecting straight lines. The four states
differ from each other in two separate domains.} \label{fig:gndsketch}
\end{figure}

Using the method described above, a ground state picture on the $H-\Delta$ plane of a particular $32^3$
realization (seed 1003) was computed and is shown in Fig.\ \ref{fig:gndstates}(a). Coexistence lines are drawn
with thickness reflecting the jump in magnetization to visualize the size of discontinuity. Most of degenerate
points are intersection points of four ground states, as illustrated in Fig.\ \ref{fig:quadr}. The four ground
states differ by the orientation of two separate domains, which are typically small, as is the discontinuity in
physical quantities between them. More interesting are triple points where three ground states are degenerate.
Here a single coexistence line bifurcates into two lines in a \textbf{Y} shape, as illustrated in Fig.\
\ref{fig:triple}. The state on the top of the \textbf{Y} results from the break-up of a relatively large domain
while this domain flips as a whole across the vertical line of the \textbf{Y}. A triple point has some
characteristics of a thermal first order transition where two ordered state coexist with a disordered state.

There are several thousand lines in the ground state picture in Fig.\ \ref{fig:allgs}, but most of these lines
have small jumps in bond energy and magnetization. We believe that only relatively large jumps contribute to the
singularity and, to emphasize these jumps, we simplify the picture by removing the lines representing small
jumps. In Fig.\ \ref{fig:simgs} is the same picture as Fig.\ \ref{fig:allgs} but a large number of lines with
small bond energy jumps ($\delta e<0.03$) have been eliminated. This simplified picture reveals a tree-like
structure built from triple points. The first order line separating the two ordered states is the trunk of the
tree, which bifurcates at the finite size critical point, located at the center of the picture, into two main
branches.  Above the finite size critical point the ground states are disordered.  The points labeled 1 and 2
correspond to the large jumps with the same labels in Fig.\ \ref{fig:ejumps}.  The inset in Fig.\
\ref{fig:simgs} shows the details of the finite size critical point and two other triple ponts immediately above
and below it.  In this paper the finite size critical point is identified as the degenerate point that maximizes
the discontinuity in the bond energy, measured by $\delta e^{*} = (|e_+ - e_0| + |e_- - e_0|)/2$, where $e_+$,
$e_-$ are the bond energies of the two ordered states, and $e_0$  is the bond energy of the disordered state,
respectively.

We propose that the critical region of the ground state picture can be rescaled in such a way that pictures for
various system sizes are statistically indistinguishable from one another.  The required scaling involves the
width of the pictured region $W_H$ in the $H$ direction, height $W_\Delta$ in the $\Delta$ direction and lower
cut-off $\delta e_{min}$ for coexistence lines retained in the picture.  The scaling of $\delta e_{min}$ should
follow Eq.\ (\ref{eq:scemin}). The picture will include a scale invariant part of the critical region if
$W_\Delta$ scales as $\Delta-\Delta_c$,
\begin{equation}
\label{eq:scwd} W_\Delta \sim L^{-1/\nu}.
\end{equation}
The scaling of $W_H$ is expected to be the same as the scaling of the external field $H$, which has been given
by Bray and Moore in their scaling theory of the RFIM~\cite{BrMo},
\begin{equation}
\label{eq:scwh} W_H \sim L^{(\alpha+\beta -2)/\nu}
\end{equation}

\begin{figure}
\begin{center}
\subfigure[L=10]{\includegraphics[width=1in]{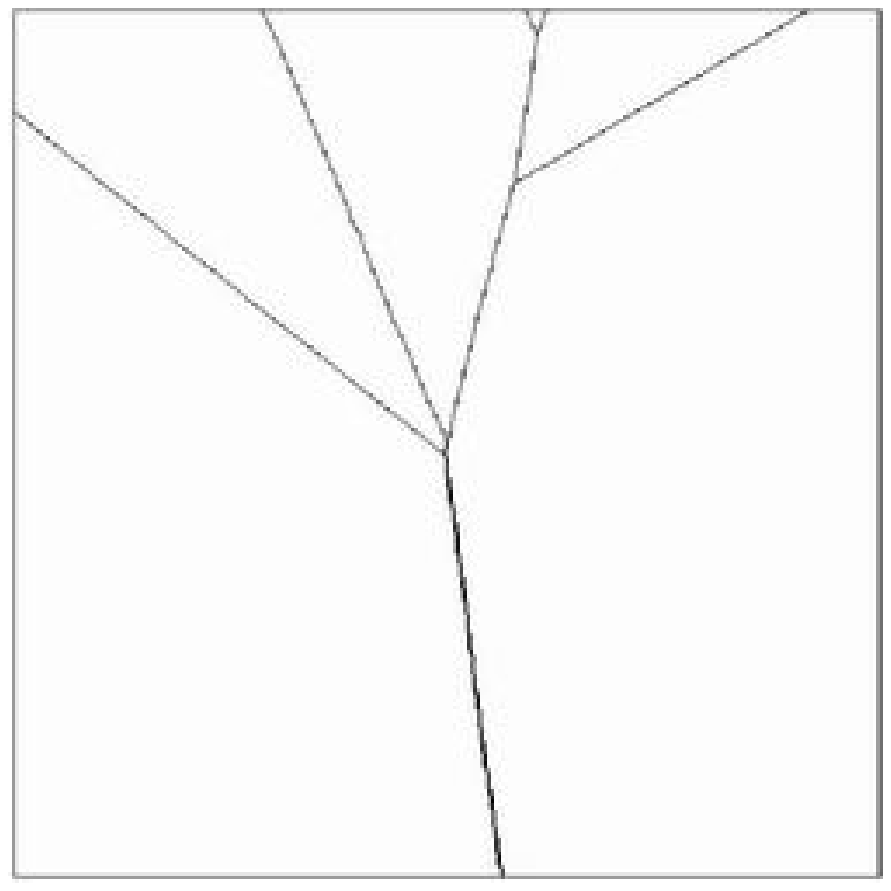}} \subfigure[L=10]{\includegraphics[width=1in]{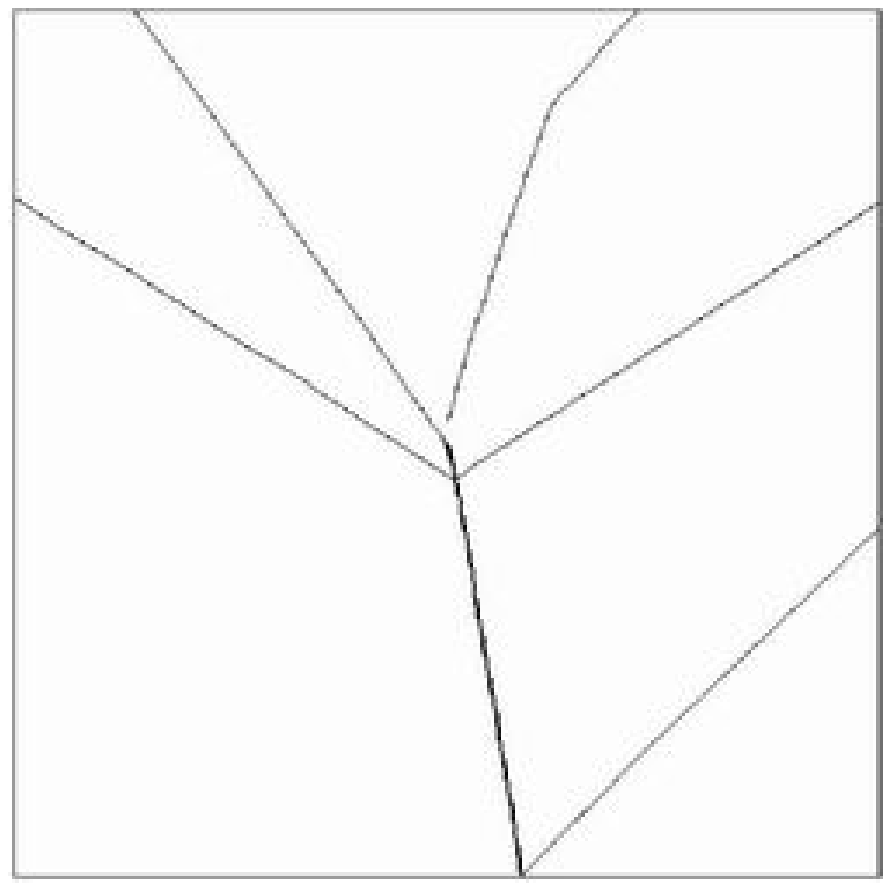}}
\subfigure[L=10]{\includegraphics[width=1in]{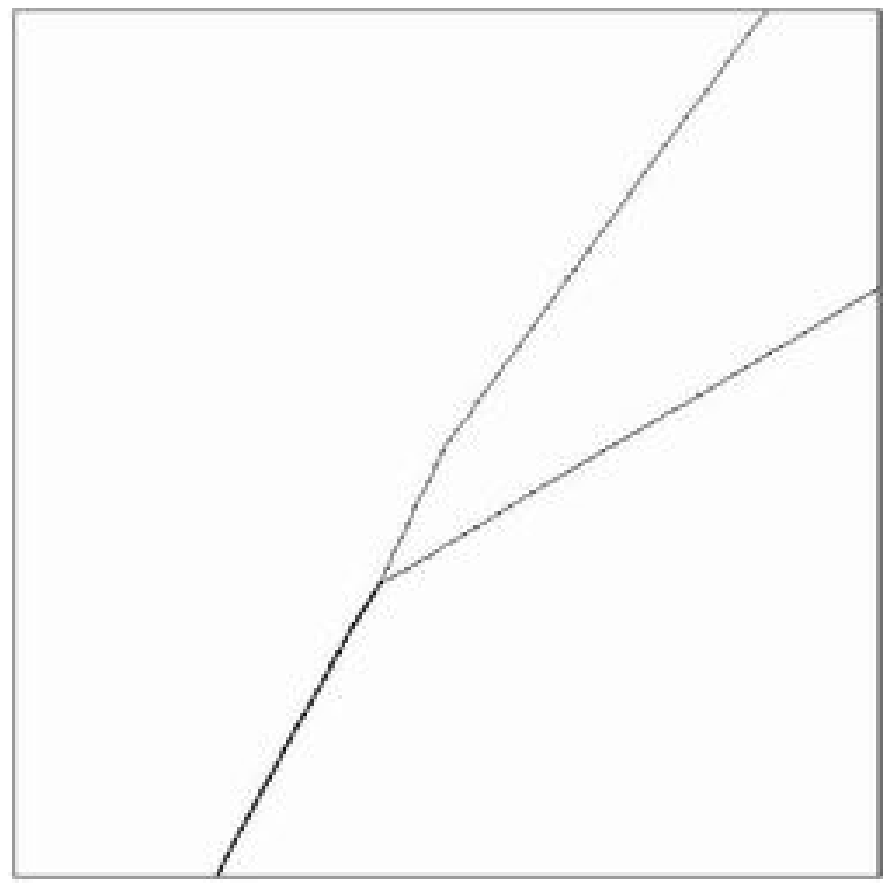}} \subfigure[L=20]{\includegraphics[width=1in]{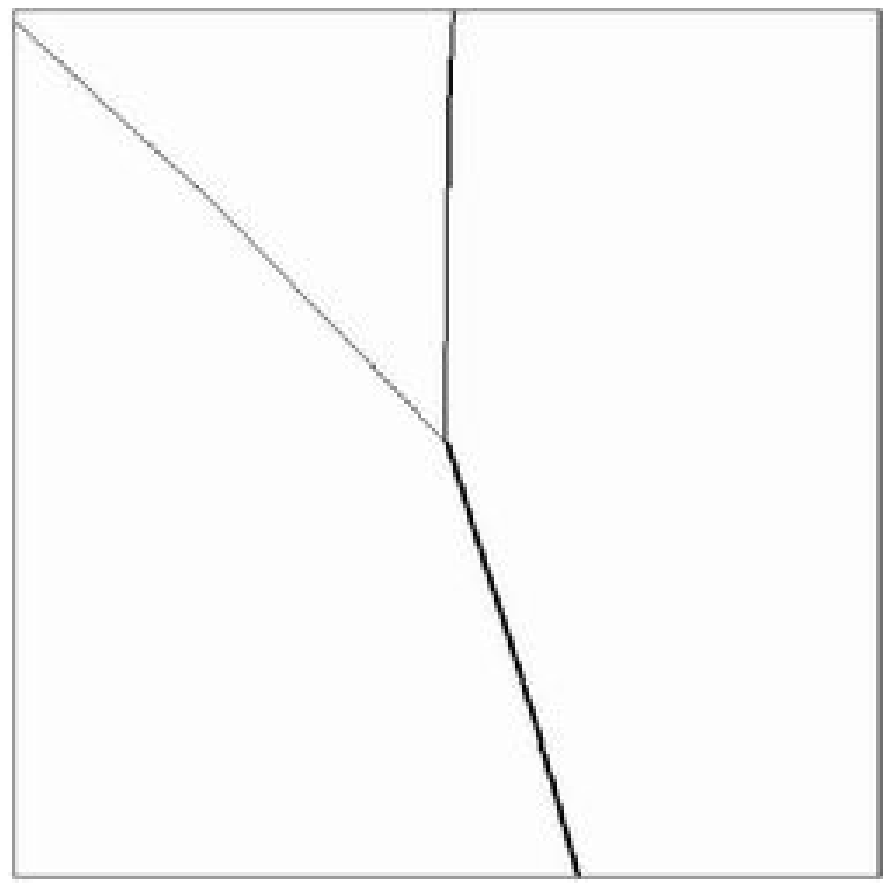}}
\subfigure[L=20]{\includegraphics[width=1in]{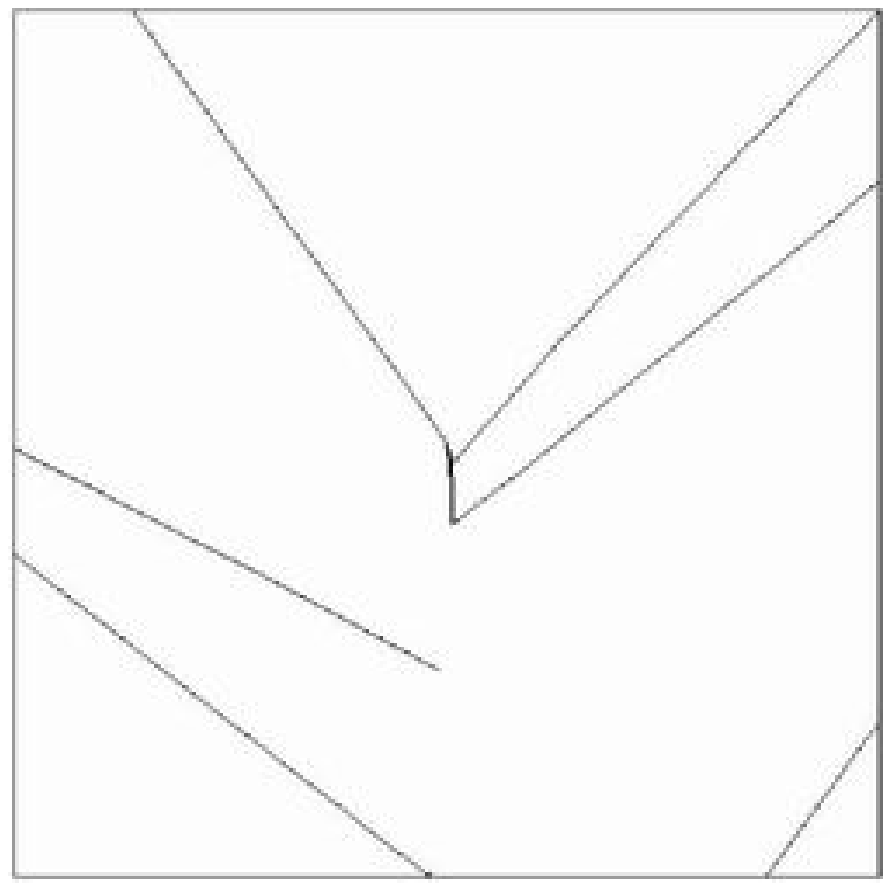}} \subfigure[L=20]{\includegraphics[width=1in]{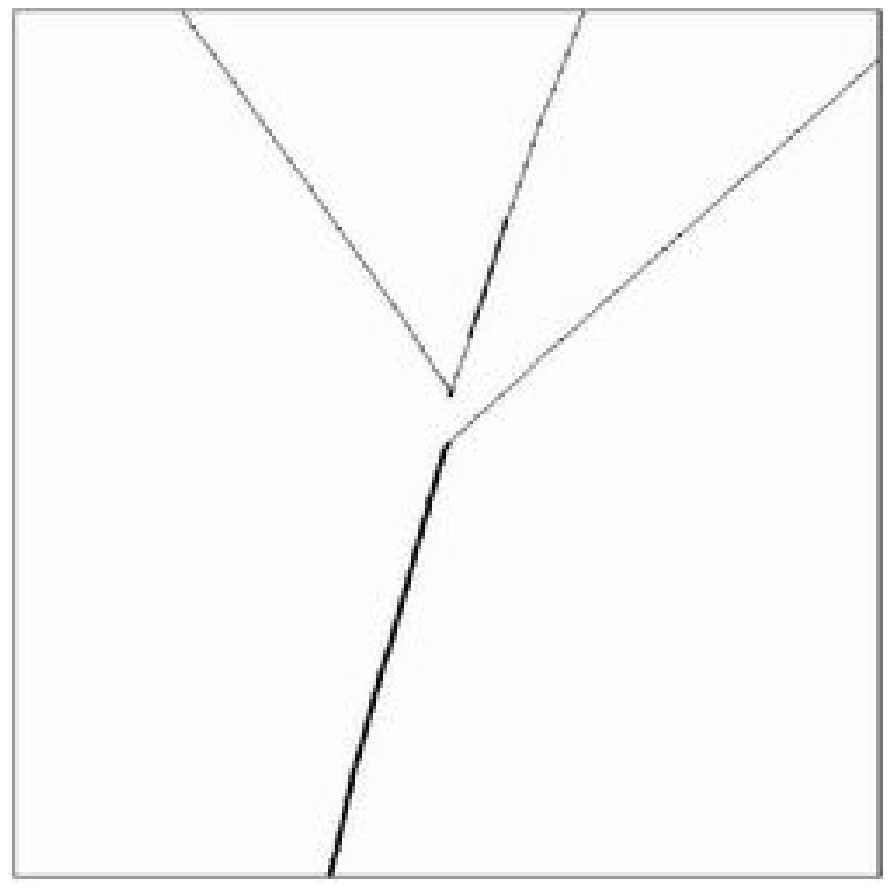}}
\subfigure[L=30]{\includegraphics[width=1in]{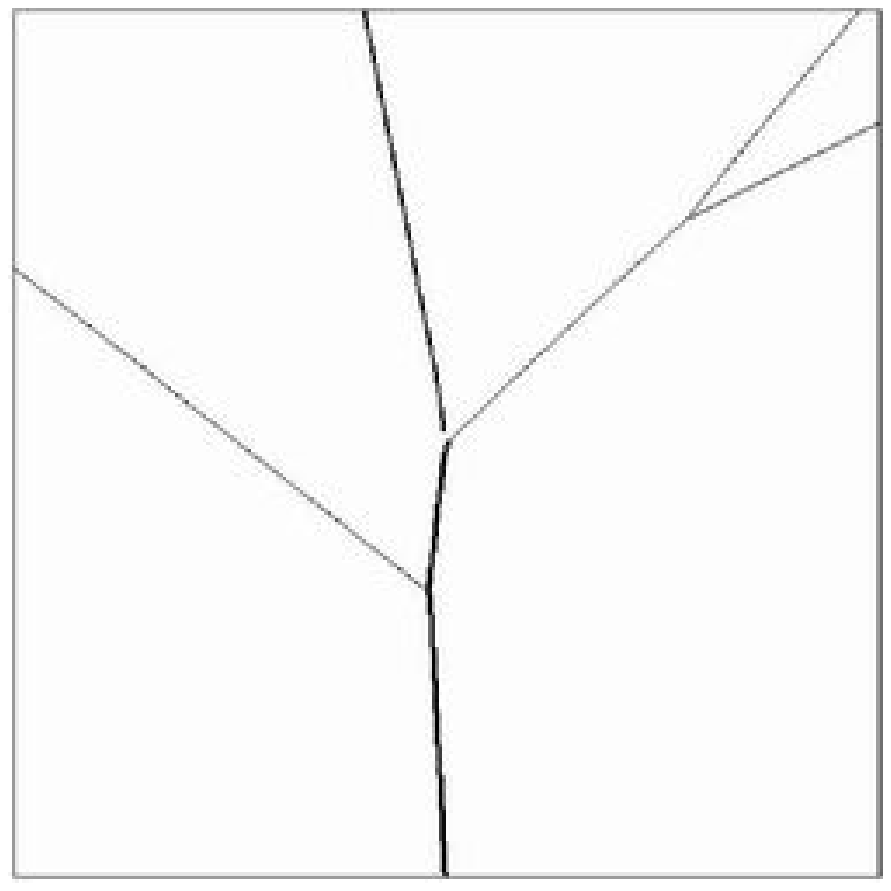}} \subfigure[L=30]{\includegraphics[width=1in]{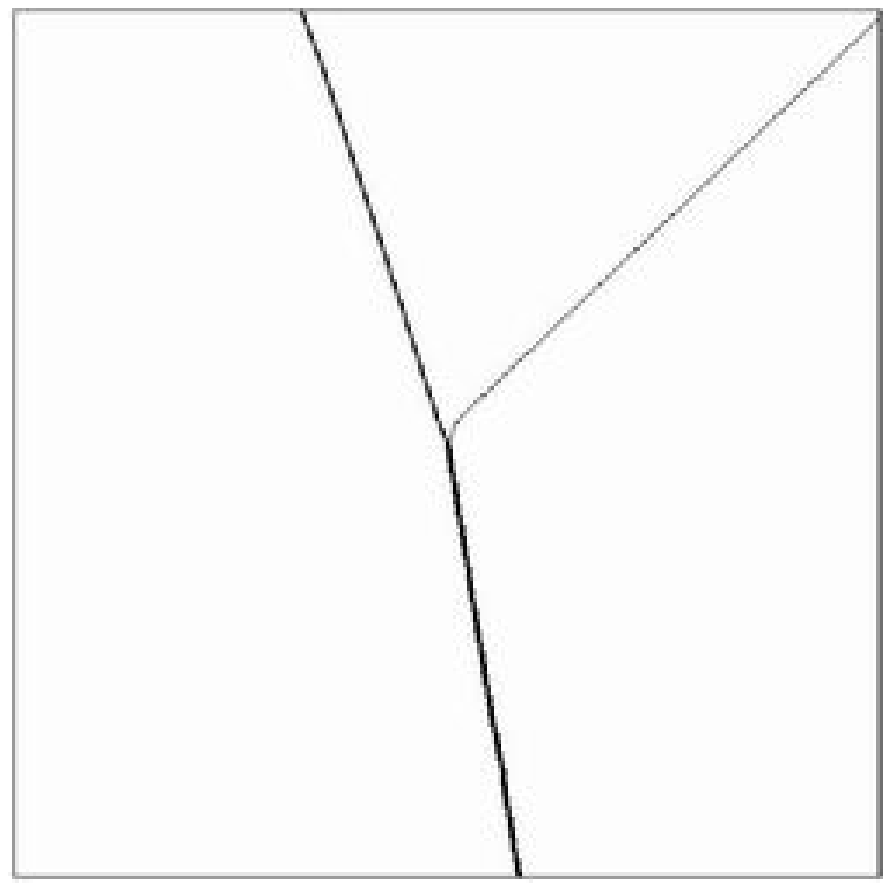}}
\subfigure[L=30]{\includegraphics[width=1in]{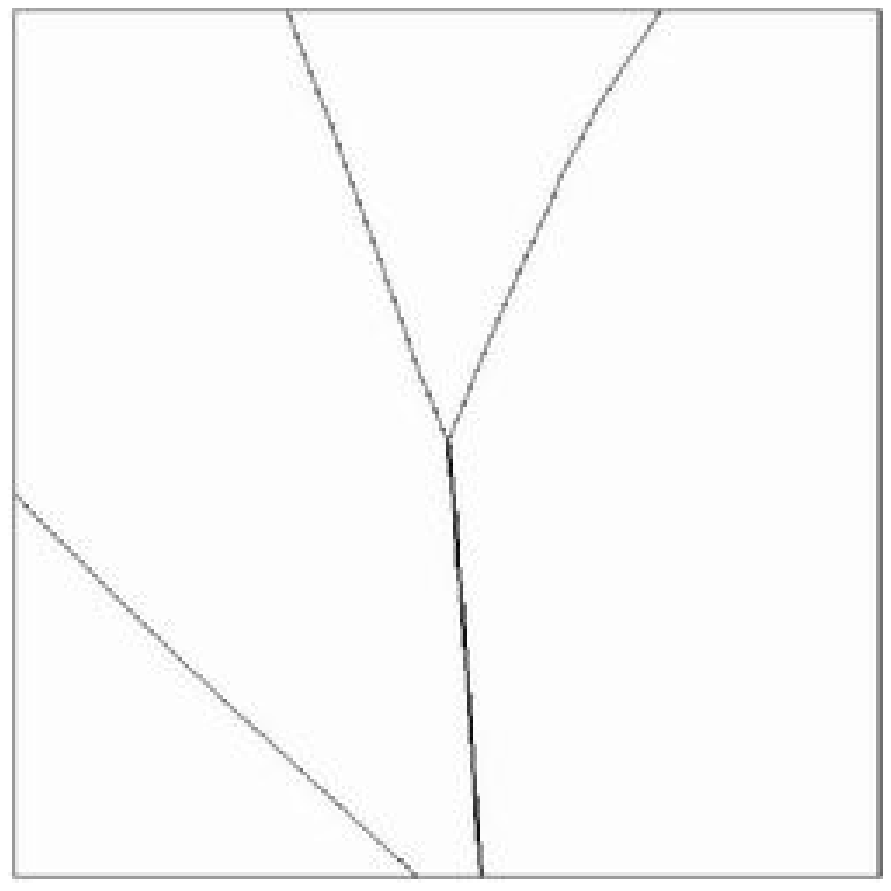}} \\
\includegraphics[scale=0.2]{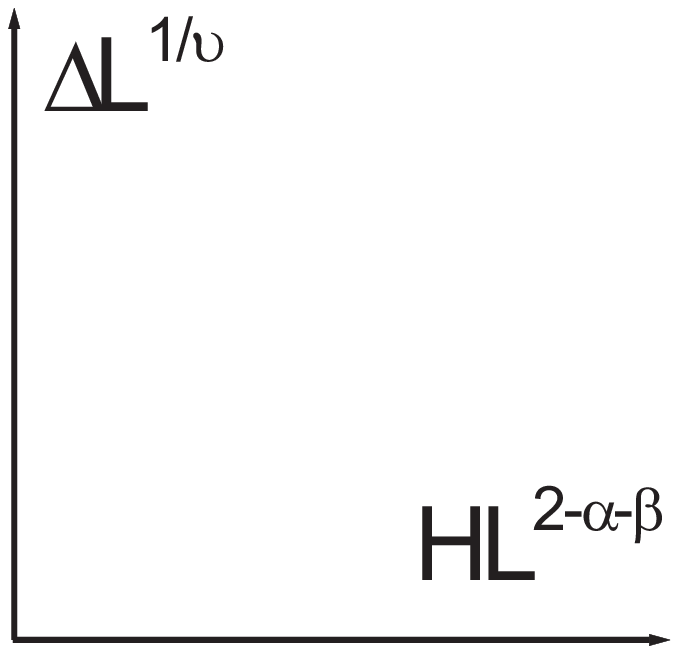}
\caption{Ground state pictures for different system sizes plotted with scaled coordinates and lower bound for
bond energy jumps. Each figure shows the scaled ground state picture for a single realization.}
\label{fig:picsc}
\end{center}
\end{figure}

In Fig.\ \ref{fig:picsc} the parameters of the pictures are scaled such that $\delta e_{min}
L^{(1-\alpha)/\nu}$, $W_\Delta L^{1/\nu}$ and $ W_H L^{(2-\alpha-\beta)/\nu}$ are all held constant. Although
different realizations have quite different ground state patterns there is no apparent way to distinguish
between different system sizes.

\begin{figure}
\includegraphics[width=3in]{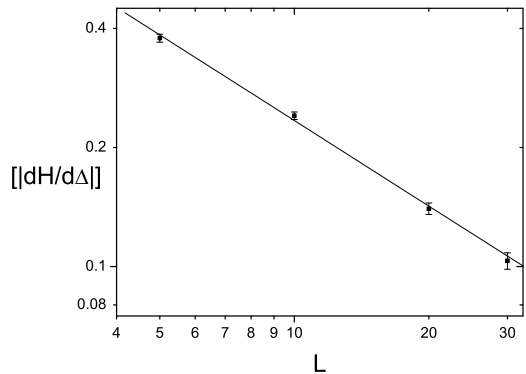}\caption{ Scaling of the average inverse slope of coexistence lines near the finite size critical point (except for the first order line). The slope of the best-fit line is $-0.79\pm0.04$, which is in agreement with the predicted value $(\alpha+\beta-1)/\nu$.}
\label{fig:slc}\end{figure}

In order to test the scaling of the ground state pictures more quantitatively we measure $\lb |{dH}/{d\Delta}|
\rb$, the average of the absolute value of the inverse slope of coexistence lines near criticality (except for
the first order line) as a function of system size. The result is shown in Fig.\ \ref{fig:slc}. We measure the
inverse slope of coexistence lines rather than the slope itself, because in some realizations the slope is very
large, and thus the average of ${d\Delta}/{dH}$ is not well-behaved. The slope of the best-fit line is
$-0.79\pm0.04$. From Eq.\ (\ref{eq:scwd}) and Eq.\ (\ref{eq:scwh}) we expect $\lb|{dH}/{d\Delta}|\rb \sim
L^{(\alpha+\beta-1)/\nu}$. The measured value $-0.79\pm0.04$ is close to $(\alpha+\beta-1)/\nu$, if
$(1-\alpha)/\nu \approx 0.84$ as we have calculated in Sec.\ \ref{sec:measure}, and $\beta\approx0$ as generally
accepted.

\begin{figure}
\includegraphics[width=3in]{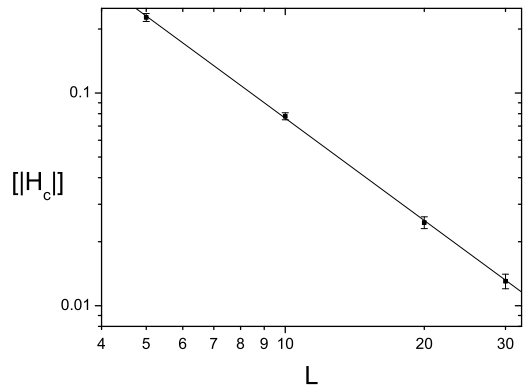}\caption{ Scaling of the average strength of the extenal field at the finite size critical point. The slope of the best-fit line is $-1.60\pm0.06$, which is in agreement with the predicted value $(\alpha+\beta-2)/\nu$.}
\label{fig:Hsc}\end{figure}

We then measure the strength of the external field at the finite size critical point $\lb |H_c| \rb$, which
should have the same scaling as $W_H$, and show the result in Fig.\ \ref{fig:Hsc}. The slope of the best-fit
line is $-1.60\pm0.06$, which is again consistent with the expected value of $(\alpha+\beta -2)/\nu$, if $\beta
\approx 0$, and our measured values of exponents $(1-\alpha)/\nu \approx 0.84$, and $1/\nu \approx 0.8$ are
used.

\section{Positive Temperature Results}
\label{sec:postemp} We have studied the RFIM at fixed disorder strength of $\Delta_0<\Delta_c$ and zero external
field for all  $T>0$ using the Wang-Landau algorithm\cite{WaLa01}. The Wang-Landau algorithm is a flat histogram
Monte Carlo method that also automatically determines the density of states $g(E)$. Thermodynamic quantities
related to energy, such as the specific heat, can then be derived from the density of states at all
temperatures. In order to get the magnetization or susceptibility, we collected joint magnetization and energy
statistics. The algorithm smooths the energy landscape and improves on the performance of the conventional
Metropolis algorithm. Using the method we can determine the specific heat and susceptibility over a broad
temperature range for systems up to size $32^3$. After we obtain the density of states, we use the Metropolis
algorithm to obtain average spin configurations for selected temperatures.

\begin{figure}[]
\centering \subfigure[$16^3$, $\Delta_0 = 2.0$, seed: 1013
]{\label{fig:c1013}\includegraphics[width=1.6in]{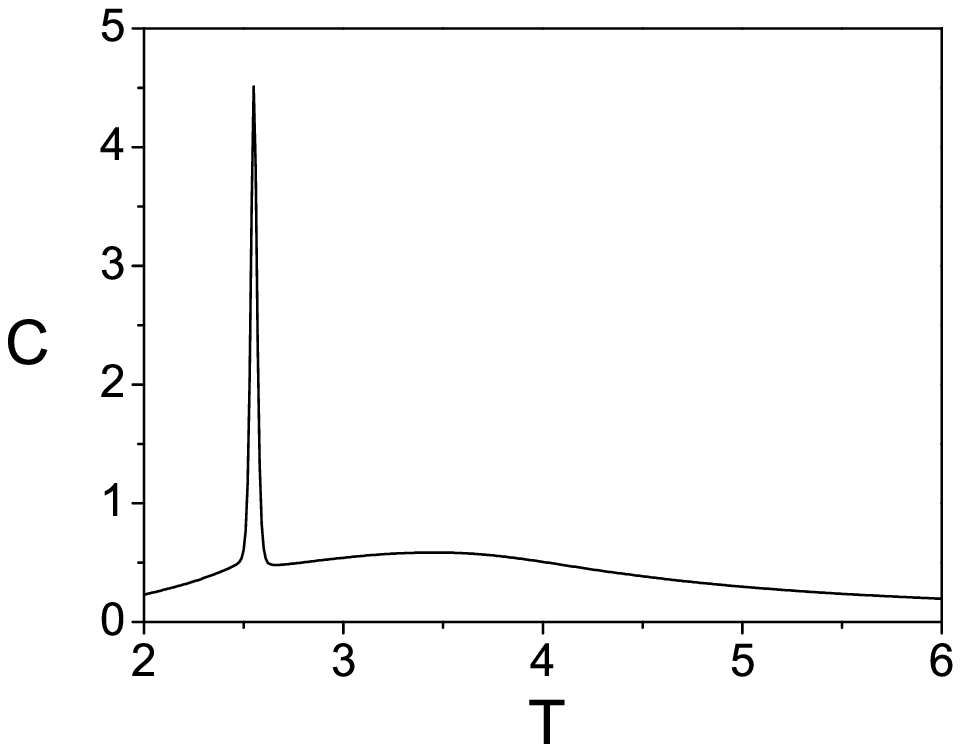}} \subfigure[$16^3$, $\Delta_0 = 2.0$, seed:
1013]{\label{fig:sus1013}\includegraphics[width=1.6in]{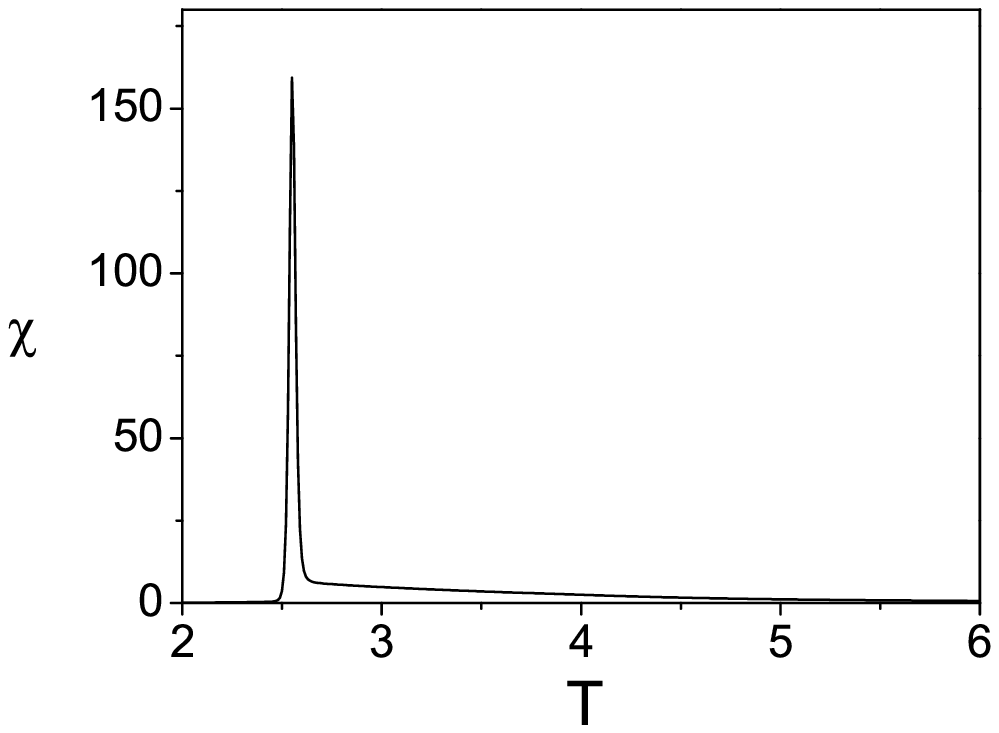}} \subfigure[$16^3$, $\Delta_0 = 2.0$,
seed: 1050]{\includegraphics[width=1.6in]{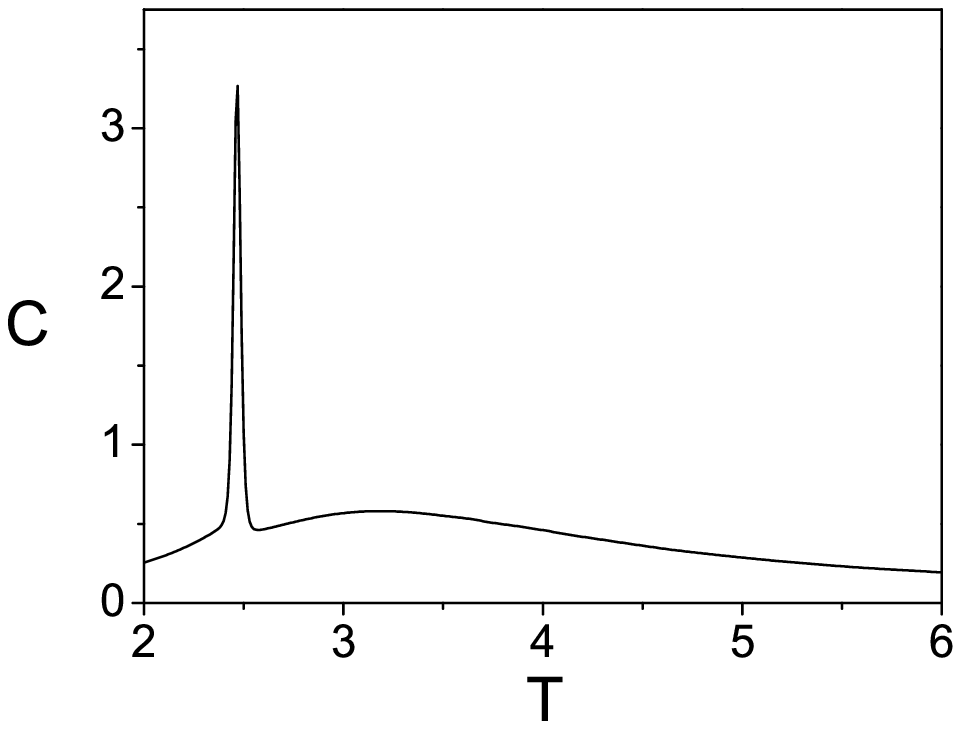}} \subfigure[$16^3$, $\Delta_0 = 2.0$, seed:
1050]{\includegraphics[width=1.6in]{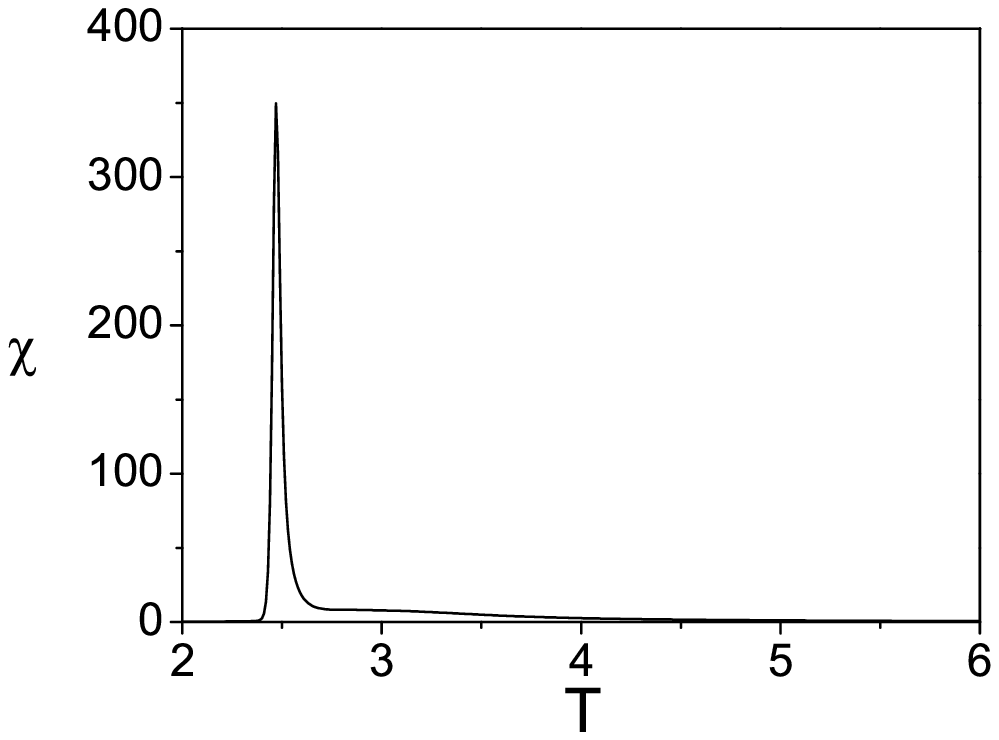}} \subfigure[$32^3$, $\Delta_0 = 1.5$, seed:
1000]{\includegraphics[width=1.6in]{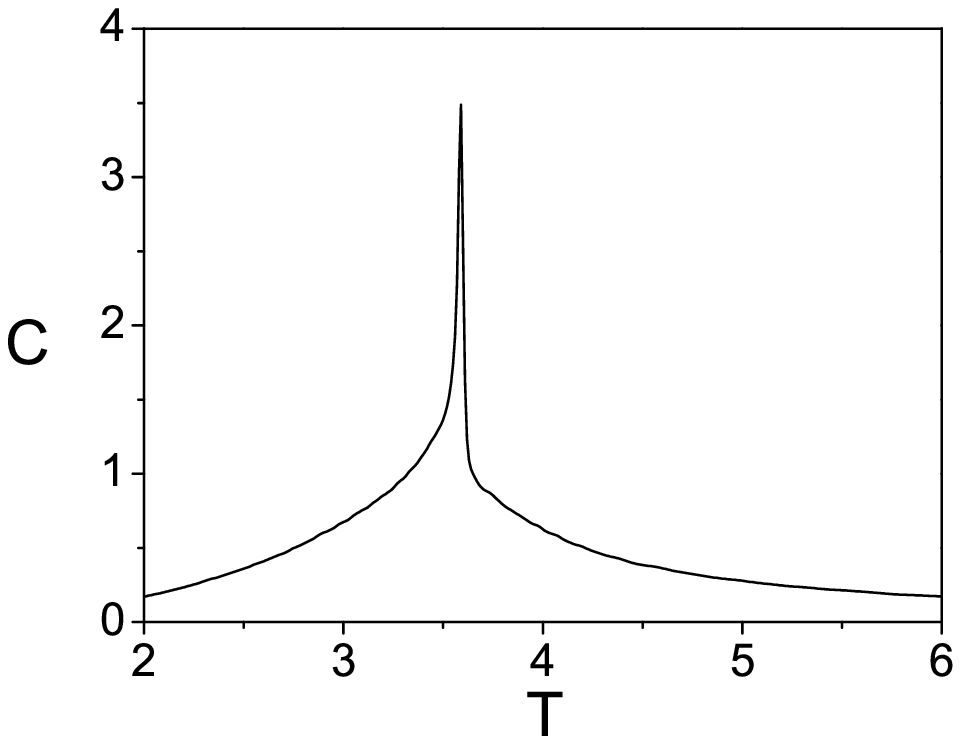}} \subfigure[$32^3$, $\Delta_0 = 1.5$, seed:
1000]{\includegraphics[width=1.6in]{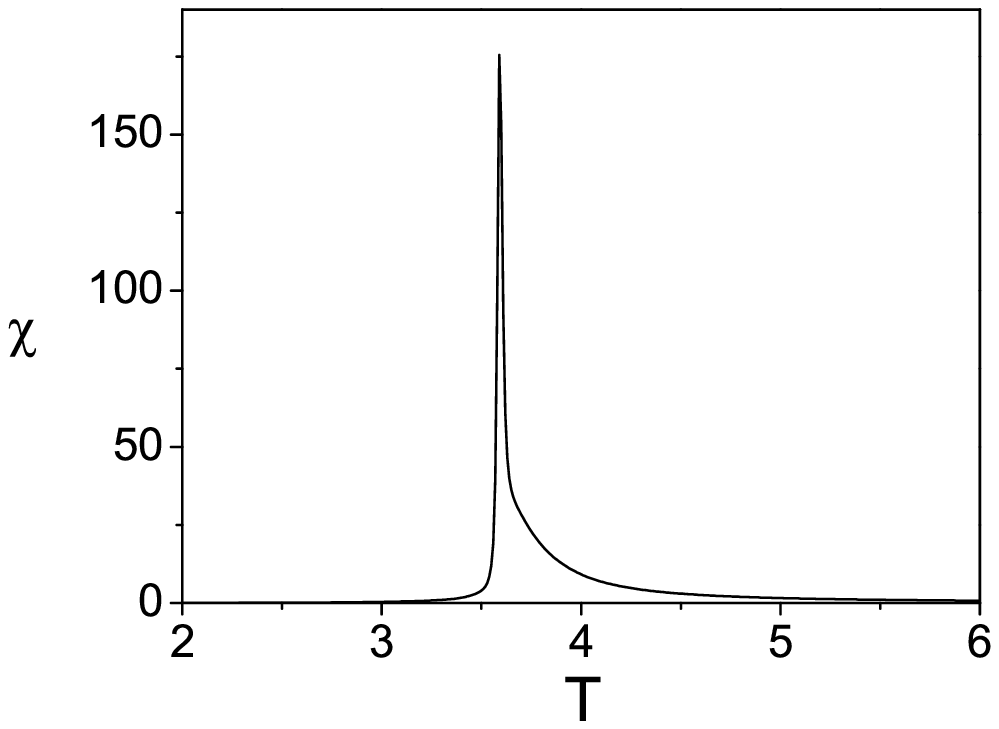}} \subfigure[$32^3$, $\Delta_0 = 2.0$, seed:
1003]{\label{fig:c1003}\includegraphics[width=1.6in]{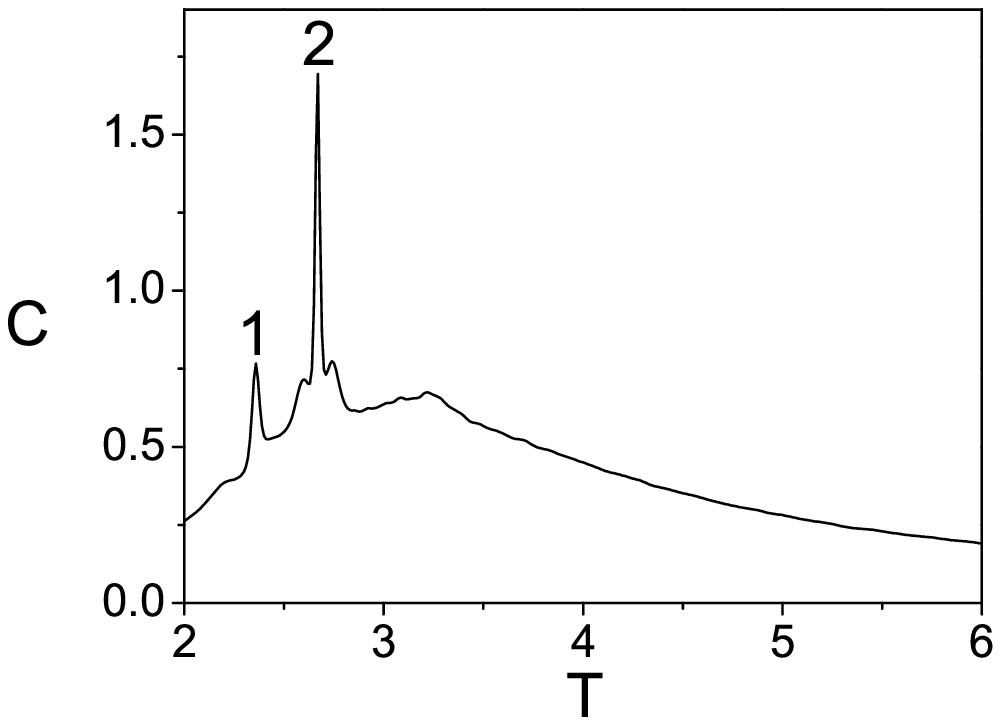}} \subfigure[$32^3$, $\Delta_0 = 2.0$, seed:
1003]{\label{fig:sus1003}\includegraphics[width=1.6in]{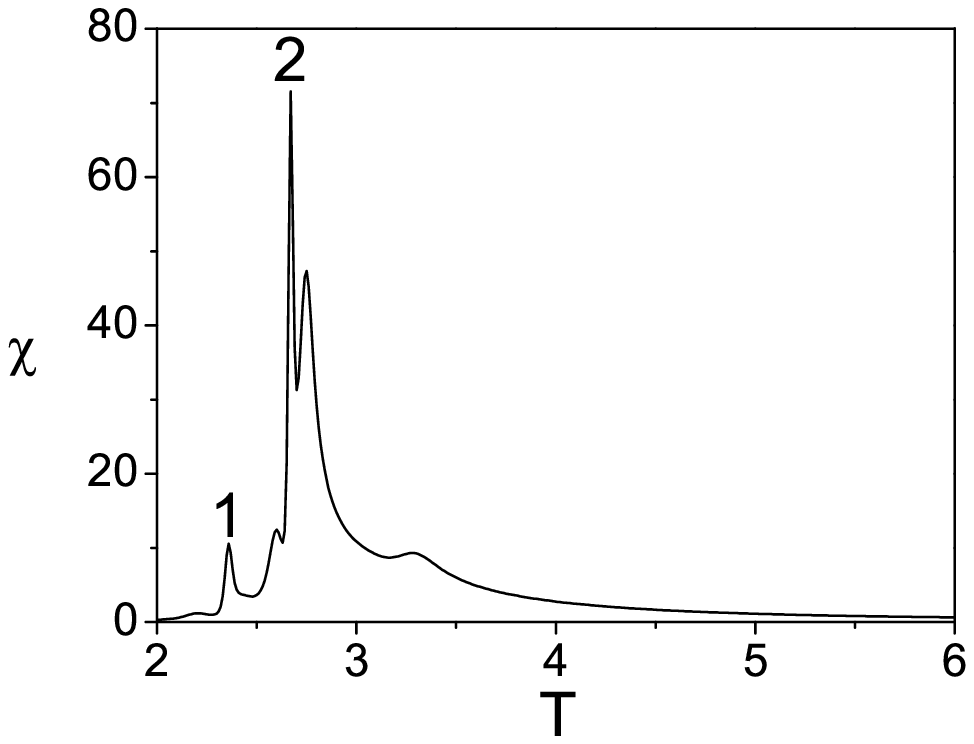}}\caption{The specific heat $C$ and the
susceptibility $\chi$ of four realizations of the RFIM. The sharp peaks in the specific heat and the
susceptibility of a given realization occur at the same temperatures.} \label{fig:sus}
\end{figure}

\begin{figure}[]
\centering \subfigure[$16^3$, $\Delta_0 = 2.0$, seed: 1013 ]{\includegraphics[width=1.6in]{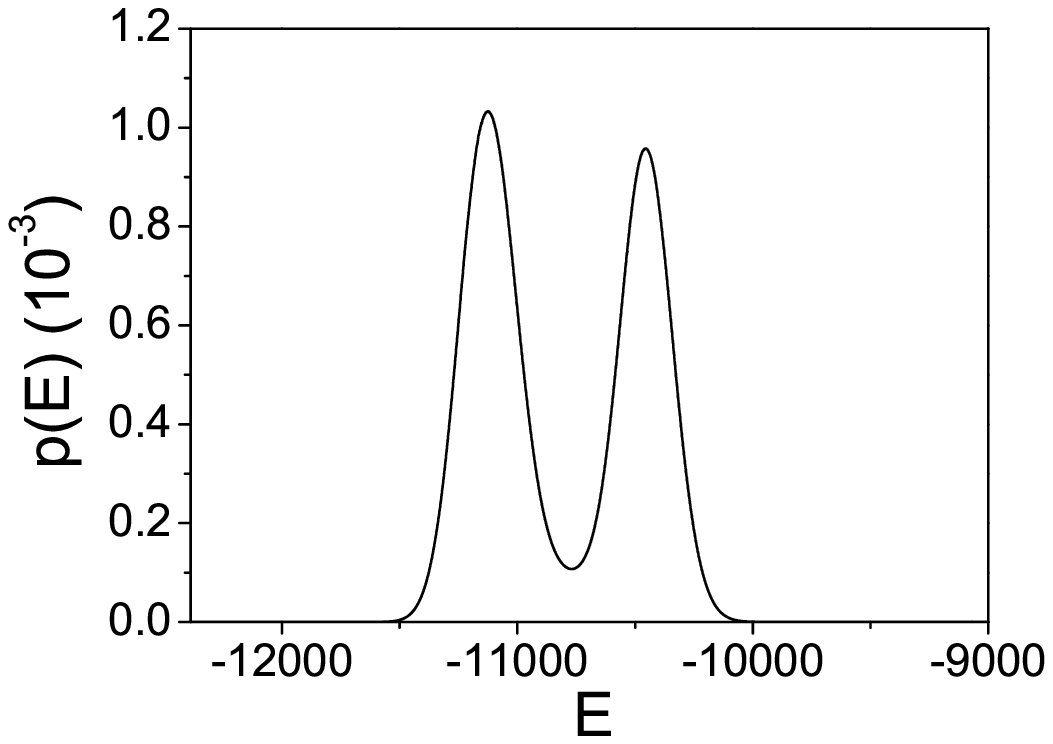}}
\subfigure[$16^3$, $\Delta_0 = 2.0$, seed: 1013]{\includegraphics[width=1.6in]{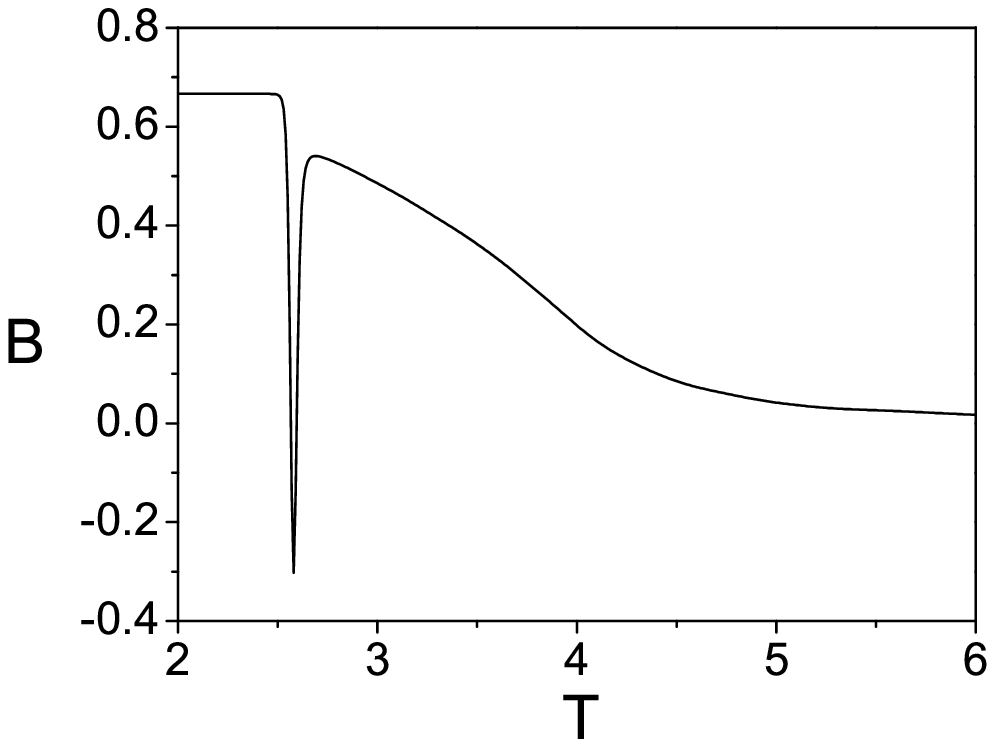}}
\caption{The first-order-like properties of sharp peaks. (a) shows the double-peaked sampling probability $p(E)$
at the sharp peak for the system in Fig.\ \ref{fig:c1013}. (b) shows the Binder cumulant $B$ as a function of
temperature for the same system, which becomes negative at the sharp-peaked transition.} \label{fig:first}
\end{figure}

Our first observation is that for some large enough systems ($\geq 16^3$) and strong enough disorder, the
specific heat and the susceptibility display one or more sharp peaks, as illustrated in Fig.\ \ref{fig:sus}. For
a given realization, the sharp peaks in the specific heat and the susceptibility occur at the same temperatures.
The sharp-peaked transitions have some first-order-like properties.  For example, the energy probability density
$p(E) = e^{-E/T} g(E) / Z$ displays double peaks, and the Binder cumulant $B(T) = 1 - \langle
m^4\rangle/3\langle m^2 \rangle$ is negative at the temperature of the sharp peaks.  The angular bracket stands
for a  thermal average. The double peaked energy distribution and negative Binder cumulant are shown for a
$16^3$ system (seed 1013) in Fig.\ \ref{fig:first}. These first-order-like features result from the coexistence
of two states that differ by flipping a large domain as we will see more clearly later. Preliminary statistics
from a small sample of realizations suggest that the fraction of realizations showing sharp peaks increases with
the system size and the strength of disorder, as shown in Table\ \ref{tab:npeak}. Here we call a transition
``sharp" if the sampling probability has two peaks at the transition temperature.

\begin{table}[]
\centering \caption{Number of realizations that have a double-peaked energy probability densities at their
specific heat peaks ($N_{dp}$), and total number of realizations simulated ($N_{tot}$) as a function of disorder
strength $\Delta_0$ and system size $L$.}
\begin{tabular*}{0.48\textwidth}{@{\extracolsep{\fill}}l c c c c}
\hline\hline $L$ & $\Delta_0$ & $N_{dp}$ & $N_{tot}$ & $N_{dp}/N_{tot}$\\
\hline
$8$ & $1.5$ &  0& 256 & 0\%\\
$8$ & $2.0$ & 0 & 64 & 0\%\\
$16$ & $1.5$ & 6 & 96 & 6.25\%\\
$16$ & $2.0$ & 21 & 96 &21.8\%\\
$32$ & $1.5$ & 3 & 9 & 33.3\%\\
$32$ & $2.0$ & 6 & 9 & 66.6\%\\
\hline\hline
\end{tabular*}
\label{tab:npeak}
\end{table}

\begin{figure}[]
\begin{center}
\includegraphics[width=3in]{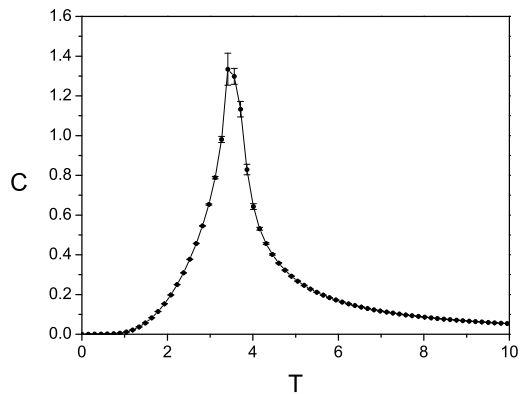}
\caption{The average specific heat of $96$ realizations of size $16^3$ and disorder $\Delta_0 = 2.0$. Although
some of these realizations have sharp peaks, the averaged specific heat is smooth.} \label{fig:avg}
\end{center}
\end{figure}

The sharp peaks occur at different temperatures with different height for different realizations, and they are
smoothed out by an average over realizations. We show in Fig.\ \ref{fig:avg} the average specific heat for all
of the $16^3$ systems we have simulated at $\Delta_0 = 2.0$. Though there are $21$ sharp-peaked realizations out
of a total of $96$ simulated (see Table\ \ref{tab:npeak}), the average specific heat is a smooth curve.  The
difference between the average specific heat and that of individual realizations shows that there is no
self-averaging at positive temperature, similar to what we have already seen at zero temperature.   The lack of
self-averaging has also been observed in the bimodal distribution RFIM~\cite{MaFy}.

\section{Relation between ground states and thermal states}
\label{sec:gndtherm}

The zero temperature fixed point picture of the RFIM phase transition predicts that the behavior in the critical
region at positive temperature is determined by the competition between couplings and random fields with thermal
fluctuations serving only to renormalize the strength of these quantities. The results presented in this section
suggest that a strong version of the zerio temperature scenario holds for individual realizations of normalized
random fields. We will show that the sharp peaks in the thermodynamic quantities can be matched one to one with
the large jumps at zero temperature. Furthermore, the spin configurations on either side of the sharp peaks can
be mapped onto the ground states on either side of the corresponding large jumps. Similar correlations between
ground states and thermal states were found in one dimension \cite{Alava}.

We illustrate the above statement for one $32^3$ realization ($\Delta_0=2.0$, seed $1003$) whose specific heat
and susceptibility are shown in Fig.\ \ref{fig:c1003} and \ref{fig:sus1003}, respectively. There are two major
peaks in the specific heat and the susceptibility, and each of them are related to the two major jumps in the
bond energy and the magnetization at zero temperature, as shown in Figs.\ \ref{fig:ejumps} and \ref{fig:simgs}
(labeled as $1$ and $2$).

The connection between the zero temperature transitions and positive temperature transitions is confirmed by the
correlation between the average spin configurations near the positive temperature transition and the ground
states near the zero temperature transition. For a single realization of random fields, we obtain the thermally
averaged spin configuration at a given temperature near the peaks using the Metropolis algorithm. We start our
simulation from the ground state, and then employ the Wang-Landau algorithm without modifying the already
obtained density of states, until the microstate falls into an energy bin that has a significant sampling
probability at the given temperature. The Metropolis updates are then used to obtain the averaged spin
configuration.

\begin{figure}
\subfigure[]{\label{fig:g1}\includegraphics[width=1in]{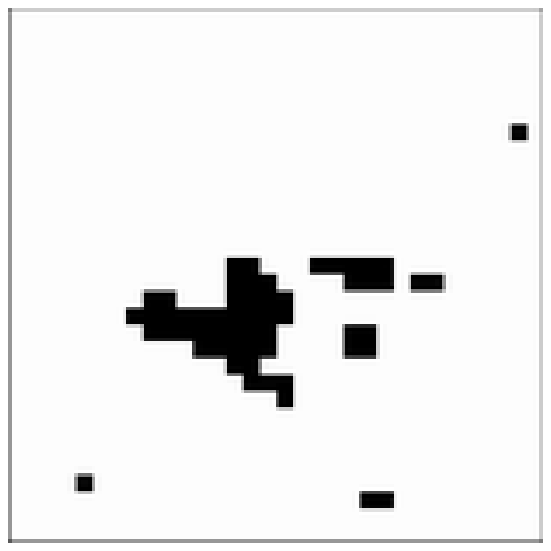}}
\subfigure[]{\label{fig:g2}\includegraphics[width=1in]{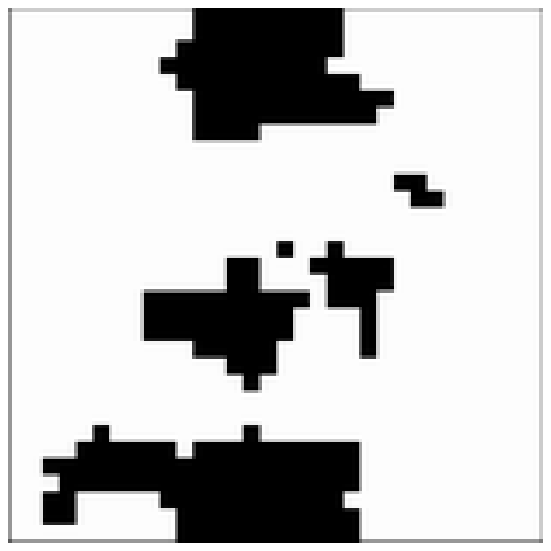}}
\subfigure[]{\label{fig:g3}\includegraphics[width=1in]{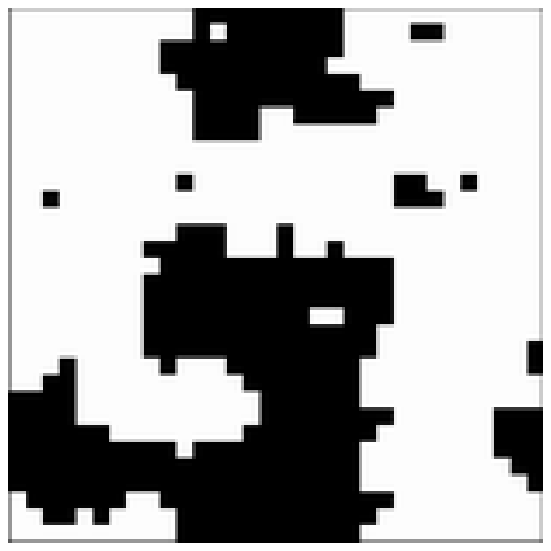}}
\subfigure[]{\label{fig:s1}\includegraphics[width=1in]{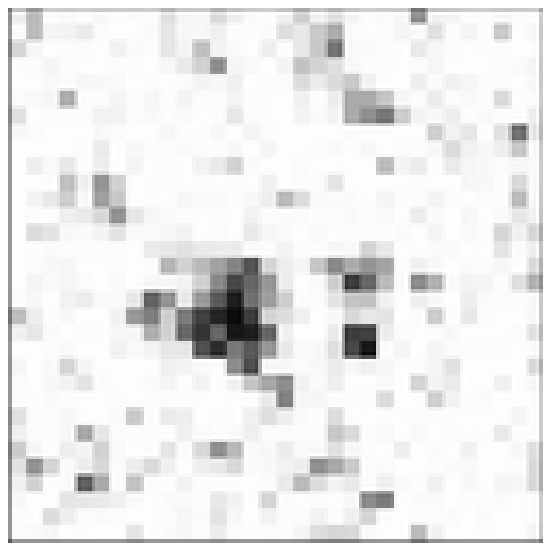}}
\subfigure[]{\label{fig:s2}\includegraphics[width=1in]{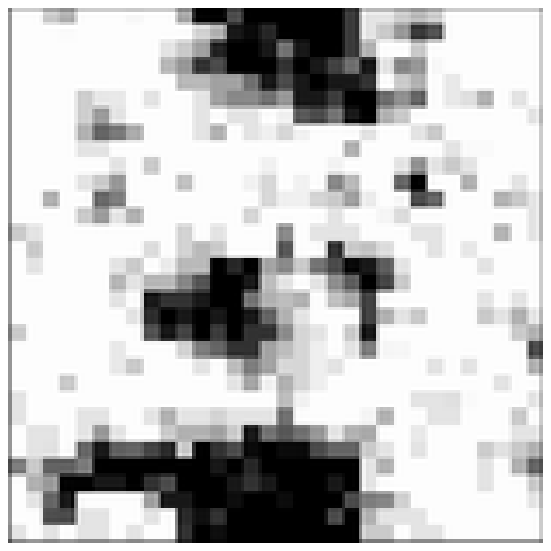}}
\subfigure[]{\label{fig:s3}\includegraphics[width=1in]{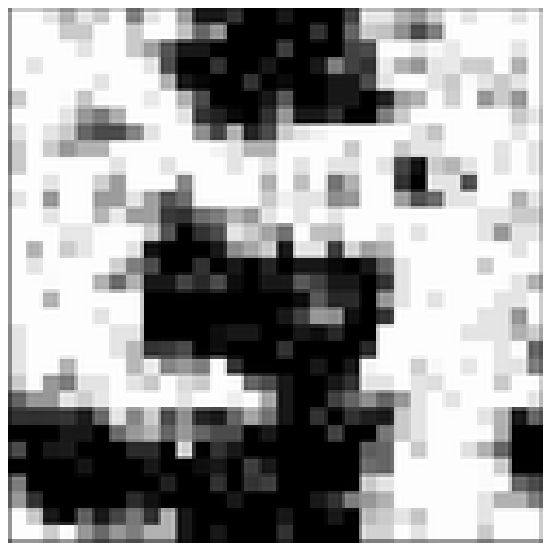}}
\subfigure[]{\label{fig:r1}\includegraphics[width=1in]{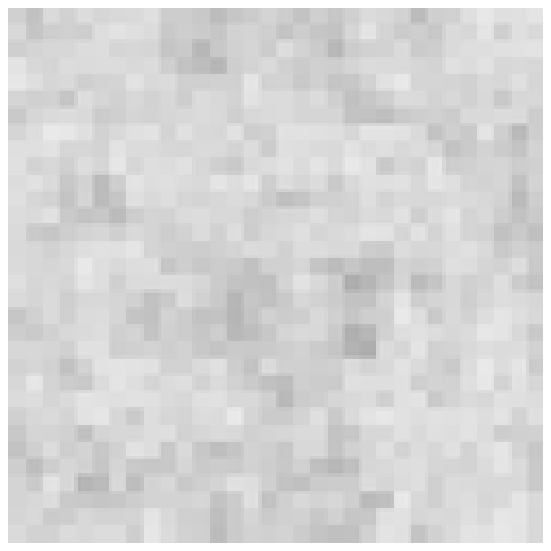}}
\subfigure[]{\label{fig:r2}\includegraphics[width=1in]{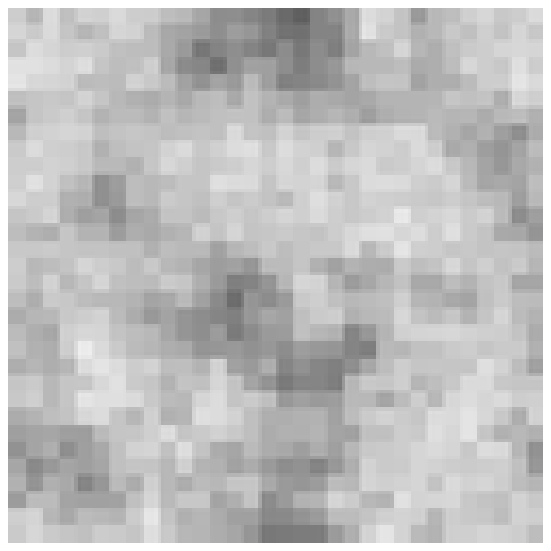}}
\subfigure[]{\label{fig:r3}\includegraphics[width=1in]{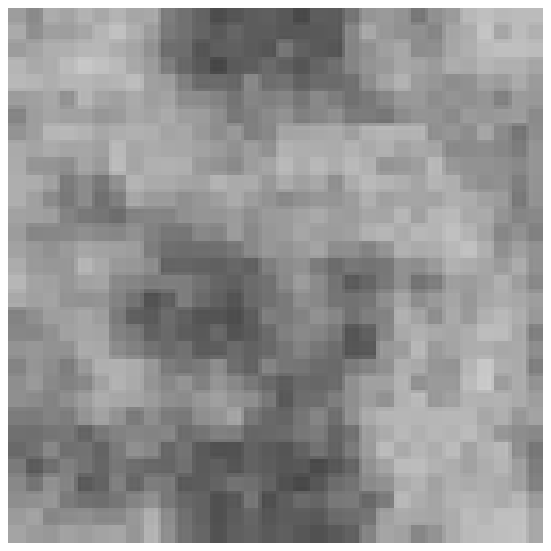}} \caption{Spin configurations near the critical
points at zero temperature and finite temperatures for a single realization of normalized random fields. Each
panel is the same plane of a $32^3$ realization with black representing spin down; white, spin up; and shades of
gray, the thermally averaged spin state.  From left to right in the top two rows, panels are at $\Delta$ ($T$)
before, between and after jumps (peaks) 1 and 2 in Fig.\ \ref{fig:simgs} (Fig.\ \ref{fig:c1003}). Specifically,
panels (a), (b) and (c) are ground states at $\Delta=2.36$, $2.41$ and $2.54$, respectively. Panels (d), (e) and
(f) are at $\Delta=2.0$ and $T=2.2$, $2.5$ and $2.8$, respectively.   Panels (g), (h) and (i) are at
$\Delta=0.5$ and temperatures 4.0, 4.3 and 4.45, near the peak in the specific heat at $T=4.375$. }
\label{fig:spin}
\end{figure}

Figures \ref{fig:s1}, \ref{fig:s2} and \ref{fig:s3} show one plane through the system with $\Delta_0 =2.0$ and
at temperatures just before peak $1$ ($T=2.2$), just after peak $1$ ($T=2.5$), and just after peak $2$
($T=2.8$), respectively. The difference between the states shows that the sharp peak corresponds to flipping a
relatively large domain. It is evident that these three states are strongly correlated with the ground state
spin configuration before the jump $1$ ($\Delta = 2.36$), just after jump $1$ ($\Delta = 2.41$), and just after
jump $2$ ($\Delta = 2.54$), as shown in Fig.\ \ref{fig:g1}, \ref{fig:g2} and \ref{fig:g3}, respectively. (The
labels of jumps and peaks are given in Figs.\ \ref{fig:ejumps}, \ref{fig:simgs} and  \ref{fig:c1003}.)

Some correlation between ground states and thermal states persists to much smaller values of $\Delta_0$ in a
regime where the thermodynamic properties no longer display sharp peaks.  Figure \ref{fig:r1}, \ref{fig:r2} and
\ref{fig:r3} show the same realization of disorder and the same plane through the system but with
$\Delta_0=0.5$.  Here the specific heat has a rounded peak at $T=4.375$. Figures \ref{fig:spin}(g), (h) and (i)
correspond to temperatures 4.0, 4.3 and 4.45, respectively.  Although there is considerable thermal ``blurring"
in these pictures, evidence of the ground states is unmistakable.

\begin{figure}
\includegraphics[width=3in]{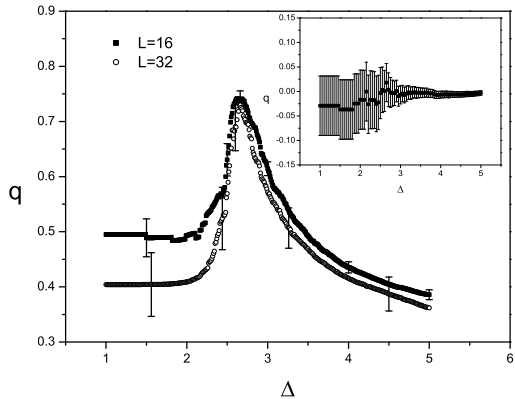}
\caption{Disorder averaged correlation $q$ of a thermal state just above the transition temperature at
$\Delta_0=1.5$ to ground states at disorder strength $\Delta$ for the same realization of random fields. Solid
squares for size $16^3$ and open circles for size $32^3$. Only a few error bars are drawn to make the figure
easier to read. The inset shows the correlation of thermal states with ground states of a different random field
realization.} \label{fig:corr}
\end{figure}

A quantitative characterization of the correlation between ground states and thermal states can be obtained from
the correlation measure,

\begin{equation}
\label{eq:corr} q(\Delta) = \frac{1}{L^3} \sum_i \, \lb \operatorname{sgn}( \langle s_i | \Delta,0 \rangle
\langle s_i | \Delta_0,T^\ast \rangle )\rb
\end{equation} where the square brackets are an average over realizations of disorder and
$\langle s_i |\Delta,T \rangle$ is the thermal average of the spin at the $i$th site at disorder $\Delta$ and
temperature $T$ or, if $T=0$, it is the ground state spin value.  For each realization,  the temperature
$T^\ast=T_{\max} + 0.1$ where $T_{\max}$ is the temperature of the maximum of the specific heat. Thus, for each
realization, we pick a thermal state just above the transition temperature. Figure \ref{fig:corr} shows $q$ vs.\
$\Delta$ for sizes $16^3$ and $32^3$ and $\Delta_0=1.5$, with $96$ realizations for size $16^3$ and $9$ for size
$32^3$.  A peak in the correlation occurs at  $\Delta \approx 2.65$ where $q \approx 0.75$.  The value, $\Delta
\approx 2.65$, is about $0.15$ larger than the average $\Delta$ at the largest discontinuity in the bond energy
for system size $32^3$. The inset in Fig.\ \ref{fig:corr} shows the average correlation between thermal states
of one realization and ground states of another for size $16^3$, which is nearly zero as expected. A second
measure, $q^\ast$ is obtained by choosing the value $\Delta^\ast$ in Eq.\ (\ref{eq:corr}) for each ground state
realization  to give  the largest correlation to the thermal state at $T^\ast$ and then averaging over
realizations.  We find that for size $32^3$, $q^\ast = 0.80 \pm 0.06 $ for $\Delta_0=1.5$ and $q^\ast = 0.85 \pm
0.05$ for $\Delta_0=2.0$.  Together, these result provide quantitative confirmation that the thermal states at
temperatures slightly above the thermal critical point are strongly correlated with the ground states at
disorder strength slightly higher than the zero temperature critical point.

The correlation between thermal states and ground states near the transition is consistent with, but not
predicted by the zero temperature fixed point hypothesis\footnote{The correlation between critical ground states
and thermal states is predicted from linear response theory, private communication M.\ Schwartz and Ref.\
\onlinecite{ScSo}.}. This hypothesis predicts that the renormalization group flow is to a zero temperature fixed
point so that the zero temperature and positive temperature transitions are in the same universality class.
However, it does not predict anything about the spin configurations along the critical line for individual
realizations of disorder. If the correlation of spin configurations along the critical line that we observe for
small systems persists to large systems, it will support the following strong version of the zero temperature
fixed point scenario: for a given realization of normalized random fields, the sequence of states near the zero
temperature critical point obtained by varying $\Delta$ for $T=0$ can be mapped onto the sequence of thermal
states near the critical point obtained by varying $T$ for fixed values of $\Delta_0$, $\Delta_0<\Delta_c$.

\section{Summary}

In this paper we have numerically studied the RFIM at zero temperature and positive temperature. At zero
temperature we have extracted critical exponents from the finite size scaling of the several largest jumps in
the bond energy. Our measured value of exponents (except $\nu$) are mostly consistent with previous values but
have better accuracy. We have found that the heat capacity exponents $\alpha$ is near zero. We have also
portrayed all ground states within a small critical region on the $H-\Delta$ plane for up to $32^3$ systems. The
ground state pictures shows a tree-like structure if small jumps are removed. Although the ground state pictures
are not self-averaging, they satisfy statistical scaling relations. That is, within a scaled region in
$H-\Delta$ plane, with scaled lower limit of bond energy jumps chosen, the ground state pictures of different
system sizes are statistically similar.

We have used the Wang-Landau algorithm to study the RFIM at positive temperature. This algorithm enabled us to
obtain the density of states and to derive the specific heat and susceptibility over a broad range of
temperatures for systems up to size $32^3$. We have observed that for some disorder realizations the transition
is characterized by sharp peaks in the specific heat and the susceptibility. The sharp-peaked transition has
some first-order-like features and the fraction of realizations that have sharp peaks increases as the system
size or the strength of disorder increases. The sharp peaks in the thermodynamic functions result from flipping
a large domain and are related to large jumps in bond energy and magnetization at zero temperature. More
specifically, the thermal average spin configurations near the finite temperature transition are correlated to
the ground states near some corresponding large jump at zero temperature. This phenomenon suggests a strong
version of the zero temperature fixed point scenario.  It remains to be seen whether the correlation between
critical ground states and thermal states persists to large systems.

\acknowledgements We acknowledge the support of NSF through grants DMR-0414122 (YW) and DMR-0242402 (YW and JM).

\end{document}